\documentclass[lettersize,journal]{IEEEtran}
\usepackage{amsmath,amsfonts}
\usepackage{algorithm}
\usepackage{array}
\usepackage[gen]{eurosym}
\usepackage{textcomp}
\usepackage{stfloats}
\usepackage{url}
\usepackage{verbatim}
\usepackage{graphicx}
\usepackage{cite}
\usepackage{multirow}
\usepackage{subcaption}
\usepackage{xcolor,colortbl}
\usepackage[table]{xcolor}
\newif\ifhighlighted
\highlightedfalse   

\ifhighlighted
  \newcommand{\rev}[1]{\textcolor{blue}{#1}}
  \newcommand{\revnote}[2]{\textbf{#1:} \textcolor{blue}{#2}}
  
\else
  \newcommand{\rev}[1]{#1}
  \newcommand{\revnote}[2]{#2}
\fi

\definecolor{Gray}{gray}{0.7}

\def\BibTeX{{\rm B\kern-.05em{\sc i\kern-.025em b}\kern-.08em
    T\kern-.1667em\lower.7ex\hbox{E}\kern-.125emX}}

\usepackage{algorithm}
\usepackage{algpseudocode}
\algdef{SE}[DOWHILE]{Do}{doWhile}{\algorithmicdo}[1]{\algorithmicwhile\ #1}%
\usepackage{balance}
\usepackage{pifont}

\newcommand{\xmark}{\ding{55}}
\begin{document}

\title{FRESCO: Fast and Reliable Edge Offloading with Reputation-based Hybrid Smart Contracts}

\author{Josip Zilic, \textit{TU Wien}, Vincenzo De Maio, \textit{TU Wien, University of Leicester}, Shashikant Ilager, \textit{University of Amsterdam}, Ivona Brandic, \textit{TU Wien}
}

\markboth{Ieee Transactions on Services Computing}%
{Shell \MakeLowercase{\textit{et al.}}: A Sample Article Using IEEEtran.cls for IEEE Journals}


\maketitle

\begin{abstract}
    Mobile devices offload latency-sensitive application tasks to edge servers to satisfy applications' Quality of Service (QoS) deadlines. Consequently, ensuring reliable offloading without QoS violations is challenging in distributed and unreliable edge environments with diverse resource and reliability levels. We propose FRESCO, a fast and reliable edge offloading framework that utilizes a blockchain-based reputation system, which enhances the reliability of offloading in the distributed edge. The distributed reputation system tracks the historical performance of edge servers, while blockchain through a consensus mechanism ensures that sensitive reputation information is secured against tampering. However, blockchain consensus typically has high latency, and therefore we employ a Hybrid Smart Contract (HSC) as a \textit{reputation state manager} that automatically computes and stores reputation securely on-chain (i.e., on the blockchain) while allowing fast offloading decisions off-chain (i.e., outside of blockchain). The \textit{offloading decision engine} uses a reputation score from HSC to derive fast offloading decisions, which are based on Satisfiability Modulo Theory (SMT). The SMT can formally guarantee a feasible solution that is valuable for latency-sensitive applications that require high reliability. With a combination of an on-chain HSC reputation state manager and an off-chain SMT decision engine, FRESCO offloads tasks to reliable servers without being hindered by blockchain consensus. In our experiment, FRESCO reduces response time by up to $7.86$ times and saves energy by up to $5.4$\% compared to all baselines while minimizing QoS violations to $0.4$\% and achieving an average decision time of just $5.05$ milliseconds.
\end{abstract}


\section{Introduction} \label{sec:introduction}
\IEEEPARstart{L}{ATENCY-SENSITIVE} mobile applications are subject to strict Quality of Service (QoS) requirements to enhance the user experience~\cite{mobiar2010,ren2021adaptive,wang2021context}. Such applications are resource-intensive, and executing them on resource-limited and battery-powered mobile devices can cause QoS violations. A typical solution to improve performance is to offload applications' tasks to edge servers~\cite{li2020qos,lin2020survey}. However, reliability is an issue for edge servers, due to (1) \textit{limited resources} that cannot compensate for unstable connections and lack of support systems (e.g., cooling and backup power~\cite{long2022mobility, aral2020learning}), and (2) \textit{volatile workloads} which yields inconsistent performance for shared multi-tenant edge environments~\cite{tuli2022pregan}. Consequently, offloading to unreliable edge servers can cause failures~\cite{wu2018performance,long2022mobility} and thus postpone or prolong offloading and potentially violate QoS. 

Estimating edge reliability is challenging due to \textit{geo-distribution}, with diverse resources and reliability levels \cite{zhao2022joint,liang2023reliability}, and \textit{mobility}, where changing environments and interacting with previously unknown servers is a norm~\cite{siriwardhana2021survey,long2022mobility}.

Many offloading works employ blockchain-enabled \cite{Ma2024BlockchainBasedTO,Shi2023DRLBasedVC,Zhou2021BlockchainbasedTS} or blockchain-based reputation systems \cite{sun2021rc,zhou2021blockchain,kang2021optimizing,iqbal2021blockchain} to achieve reliable offloading. As a trustworthiness metric, reputation scores can be associated with edge servers based on past performance and stored in a distributed database for informative offloading decision-making. Integrating a reputation with blockchain is sensible in malicious environments where different actors can tamper with a reputation to unjustifiably inflate selected servers while downgrading others to gain incentives unfairly, potentially leading to end-user QoS violations\cite{sun2021rc,zhou2021blockchain,kang2021optimizing,iqbal2021blockchain}. However, there are a few challenges of edge offloading with blockchain-based reputation systems remain unresolved, such as challenge \textbf{(C1)} without providing \textit{formal guarantee} about the feasibility of offloading decisions which is important for applications that require high reliability, like latency-sensitive ones, \textbf{(C2)} encountering edge servers in \textit{distributed edge} environments that have diverse reliability levels and with whom did not have any prior experience, \textbf{(C3)} not addressing reliability in terms of \textit{edge failures}, and \textbf{(C4)} neglecting the impact of long-latency blockchain consensus on offloading decisions, especially for \textit{latency-sensitive} applications.


To address these challenges, we introduce a FRESCO edge offloading framework, which optimizes both offloading and reliability in distributed unreliable edge scenarios for latency-sensitive applications. FRESCO consists of offloading and reliability components. Regarding offloading, we employ an \textit{offloading decision engine} based on a formal method called satisfiability modulo theory (SMT) which is deployed on the mobile device. The SMT addresses challenge \textbf{(C1)} by providing formal proof assurance that relevant resource limitations and timing constraints are satisfied, which fits resource-constrained and latency-sensitive settings. SMT relies on constraints and logic rather than environmental variables like heuristics or machine learning, which makes it an environment-agnostic approach, suitable for distributed scenarios \textbf(C2). 

Concerning reliability, to address challenge \textbf{(C3)}, thanks to the performance-tracking feature, the blockchain-based reputation system is re-purposed for estimating the reliability levels in terms of failures rather than trustworthiness. However, blockchain-based systems are slowly responsive due to consensus protocols that conflict with our latency objective. To enable a blockchain-based reputation system for latency-sensitive applications \textbf{(C4)}, we employ a hybrid smart contract (HSC) as a \textit{reputation state manager}. HSC allows off-chain (i.e., outside of blockchain) transactions like fast offloading decisions that require performance while retaining secured on-chain storage (i.e., on the blockchain) of sensitive reputation information against malicious tampering. The HSC, deployed on the blockchain, is queried by mobile devices to retrieve secured reputation scores about servers' reliability levels to the SMT-based decision engine for reliable offloading decisions. In summary, FRESCO bypasses slow consensus with HSC for fast off-chain offloading decisions on reliable edge servers while preserving a secured on-chain reputation.

We evaluated FRESCO against Skype availability traces, simulated latency-sensitive applications, dynamic queuing workload model, and scalable infrastructure from the OpenCellID dataset. FRESCO reduces response time by up to 7.86 times and saves energy by up to 5.4\% compared to all baselines, while minimizing QoS violations to 0.4\% and achieving a low-latency average decision time of 5.05 milliseconds.

The rest of the paper is structured as follows. Section \ref{sec:background} presents methodologies, while Section \ref{sec:model} presents the system model. Section \ref{sec:opt-problem} formalizes the problem and presents our algorithm. In Section \ref{sec:experiment}, we present our experiment and evaluation results. In Section \ref{sec:discussion}, we discuss the assumptions and limitations of our approach. Finally, the related work and conclusion are in Sections \ref{sec:related} and \ref{sec:conclusion}.

\section{Motivation and Background} \label{sec:background}
\subsection{Motivational use case}

\begin{figure}[!t]
    \centering
    \includegraphics[width=\columnwidth]{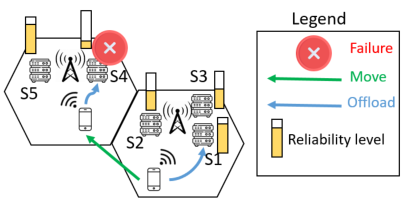}
    \caption{\rev{Reliability-aware mobile edge offloading.}}
    \label{fig:naviar-offload}
\end{figure}

\revnote{R2\_C1}{Application domains such as mobile augmented reality (MAR) are resource-intensive and latency-sensitive, requiring an abundance of resources to deliver near-instant response and ensure a good user experience. Any significant delay during MAR execution hinders the users' experience~\cite{ren2021adaptive,wang2021context}. Additionally, it exhibits high-mobility patterns and a local, limited view of surrounding resources, which makes it prone to volatile performance. An example of a MAR application instance is NaviAR, which supports users with real-time navigation by displaying virtual path information over the physical environment. An application consists of multiple interdependent tasks, viewed as a directed acyclic graph (DAG) that encodes its modularity and execution order. Tasks are offloadable, except those dependent on local functions (e.g., camera).}

\rev{NaviAR latency-sensitive tasks (e.g., calculating the shortest path) require offloading to the nearby edge servers to respect the latency constraints~\cite{wang2021context}. The Figure~\ref{fig:naviar-offload} shows a mobile device that offloads NaviAR tasks in the current cell to a server $S1$ based on its reliability level, derived from historical interactions. However, due to mobility and limited communication range, the device moves to a neighboring cell where it interacts with a different environment. In a new cell, the server $S4$ is more prone to failures and has inconsistent performance that leads to frequent deadline violations, which can cause additional delay~\cite{liang2023reliability}. Thus, without shared reliability information, the device can potentially offload to the unreliable $S4$ server instead of the reliable $S5$ server. Moreover, storing this critical aggregated reliability information from multiple devices can be prone to malicious tampering, where attackers can manipulate information to forge performance records and nudge device selection towards less reliable servers\cite{sun2021rc,yu2020crowdr,kang2021optimizing,iqbal2021blockchain}. Therefore, based on the aforementioned needs, we require latency-sensitive offloading on reliable edge servers that satisfy performance constraints based on accessible and tamper-resistant historical performance records without relying just on local information or nearby devices that can often lead to other issues like collusion\cite{deng2020trust}.}

\subsection{Blockchain-based Reputation Systems and Hybrid Smart Contracts}
Blockchain is a decentralized network that secures transactions through consensus, where all participating nodes agree on the current blockchain state. Tampering with the transactions requires owning the majority of nodes on a large-scale public blockchain (e.g., Ethereum).

A smart contract is a self-executing program that automatically enforces agreed-upon rules when certain events or conditions on the blockchain are met. Thanks to the tamper-resistance property of the blockchain, smart contracts can securely execute transactions that include sensitive information. However, the blockchain imposes long latencies and limits functionalities that smart contracts can provide by excluding non-deterministic operations (e.g., floating-point arithmetic)\cite{battah2021blockchain}. Additionally, blockchain is self-contained and accepts only transactions that occur on-chain, which makes it unsuitable for complex and latency-sensitive applications.

Hybrid smart contracts (HSC), on the other hand, allow transactions to happen off-chain, avoiding consensus overhead. This enables to perform performance-critical offloading decisions off-chain, while securing on-chain the sensitive reputation information. For instance, reputation information, as a subjective belief about the consistency of past performance, can be considered sensitive information to identify reliable servers for executing tasks. Reputation can be maliciously tampered with to take a competitive advantage over other servers, which can lead to less efficient and reliable offloading and potentially to QoS violations. Therefore, a trust-sensitive reputation system is deployed on-chain as an HSC while latency-sensitive offloading is performed off-chain.

\section{System Model} \label{sec:model}
\begin{figure}[h]
    \centering    
    \includegraphics[width=\columnwidth]{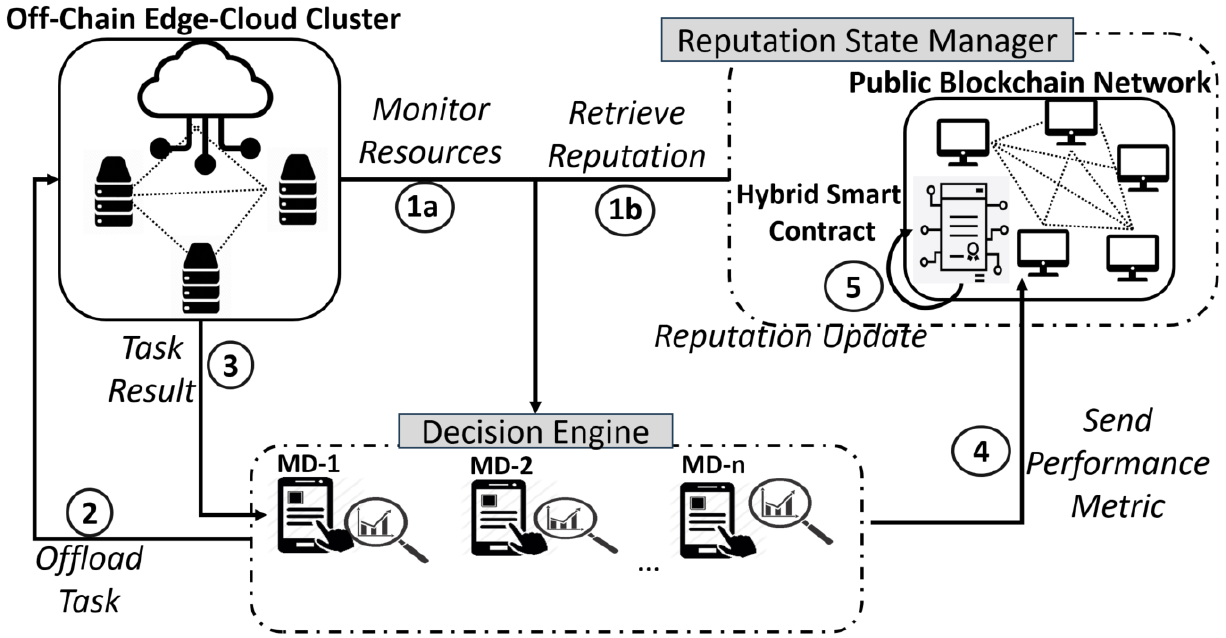}
    \caption{Edge offloading lifecycle model.}
    \label{fig:off-cycle}
\end{figure}

Figure~\ref{fig:off-cycle} illustrates the edge offloading lifecycle model, which manages offloaded tasks and estimates the reliability level of edge servers based on monitored performance. The two main components of our solution are the \textit{reputation state manager} and the \textit{offloading decision engine}. The reputation state manager is deployed as an HSC on the public blockchain network, estimates the critical reliability level of edge servers as a reputation score, and stores it securely on a public blockchain thanks to a consensus mechanism. The decision engine offloads tasks to an off-chain cluster based on reputation scores retrieved from the reputation state manager. The decision engine is often exposed as an intermediate central third-party service~\cite{lin2020survey}, making it vulnerable as a single point of failure in an unreliable environment. In our system, the decision engine is deployed on the mobile device, therefore its design choices should ensure a limited overhead, to guarantee fast decision time even on limited-resource mobile devices.

The offloading lifecycle is as follows. In steps 1a and 1b, the mobile device retrieves the reputation score from HSC and monitors resources on the off-chain cluster. Based on the procured information, the mobile device calculates offloading decisions and offloads tasks to the off-chain cluster in step 2. Task results are returned to the mobile device after execution in step 3. The mobile device records the performance metric (e.g., response time) and sends it to the HSC on the blockchain for evaluation in step 4. Finally, HSC compares the received performance metric to the deadlines and updates the reputation score accordingly. The lifecycle is repeated until the application is terminated. Noteworthy to mention, is that blockchain consensus is triggered only upon reputation update but not at reputation retrieval, which makes cached reputation score accessible in (near-)real-time.

\subsection{Queuing and response time} \label{sec:edge-queue-model}
\begin{figure}[h]
    \centering    
    \includegraphics[width=\columnwidth]{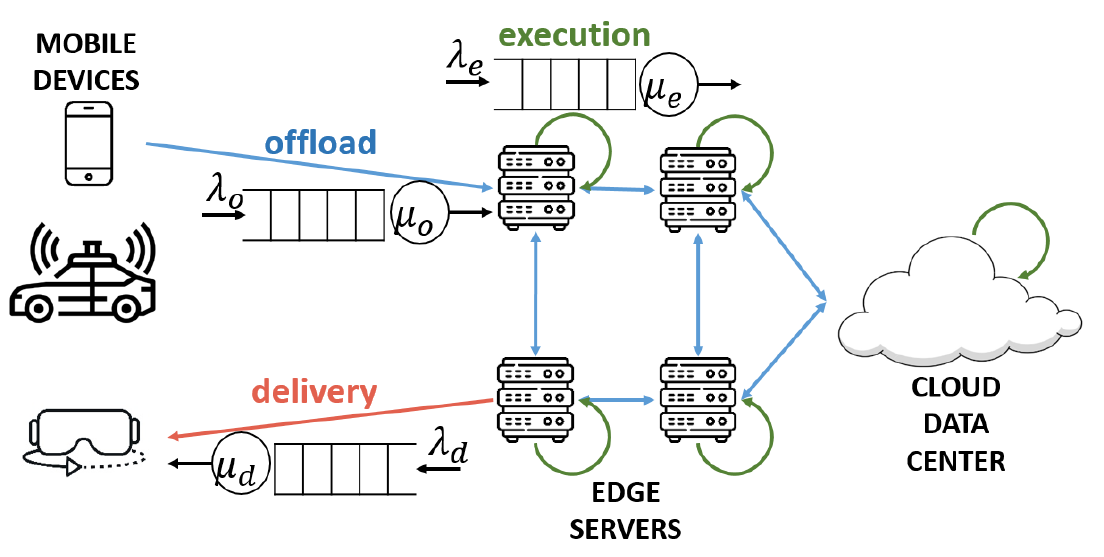}
    \caption{Dynamic queuing workload model}
    \label{fig:sys-queue-model}
\end{figure}

The workload on the shared infrastructure can be highly dynamic, where response times are hard to predict due to heterogeneous resources and tasks. To describe such dynamic behavior, we employ a queuing theory. Figure~\ref{fig:sys-queue-model} illustrates the dynamic queuing workload model at the edge, which consists of three queuing systems. The \emph{task offloading queue} models the task offloading, where multiple mobile task sources generate and offload tasks to remote servers through a shared communication channel. The \emph{task execution queue} model the execution of the task, where the servers share their resources to execute multiple tasks. The \emph{task result delivery} queue models the delivery of task results, where the results are sent back to the sources through the shared channel. The response time is defined as $RT(v,t,h) = T_{o}(h,t) + T_{e}(v,t) + T_{d}(h,t)$ where $t$ is a task, $h$ is a communication channel between pairs that can correspond to devices and servers, $v \in V$ where $V = N \cup \{m\}$ is a set of task execution nodes where a task can be executed on remote edge and cloud servers $N$ and local mobile device $m$, $T_{o}(h,t)$ is offloading latency, $T_{e}(v,t)$ is execution latency and $T_{d}(h,t)$ is delivery latency. We assume that the channels are on distinct frequencies to avoid interference \cite{ansari2018workload}, and we employ a nonpreemptive First-Come-First-Served queuing policy that makes the performance predictable, which is important for high-reliability applications.


\subsubsection{Communication latency}
The shared channels in offloading and delivery are modeled as M/M/1 queues, which emulates practical transmission due to fair sharing and different bandwidths \cite{ansari2018workload}. We will use the term communication and symbol $T_{c}(h,t)$ when referring to offloading and delivery.

The arrival process relates to task generation, modeled as Poisson with an arrival rate $\lambda_{c}(s)$ where $s \in S$ is a node in the task load generation set $S = G \cup N \cup \{m\}$ which consists of mobile devices $G$ that have the sole function of generating tasks, remote servers $N$ and mobile device $m$ which has the decision engine deployed. Task sizes are sampled from the exponential distribution with task size rate $data(t)$, accounting for the diversity of tasks' $t$ resources. Generated tasks occupy shared resources, where bandwidth utilization $U_{c}(h,t)$ is defined as a ratio of generated tasks $\lambda_c(s) \cdot data(t)$ and total bandwidth $bw_{total}(h)$:

\begin{equation} \label{eq:utilization}
    U_{c}(h,t) = \sum_{s \in D_c} \frac{\lambda_c(s) \cdot data(t)}{bw_{total}(h)}
\end{equation}

where $D_c\subset S$ represents a subset of load generators in a specific communication direction, like $D_o=G\cup\{m\}$ represents task generators and mobile devices that generate tasks in the offloading channel $c=o$, or only remote servers $D_d=N$ on a delivery channel when task results are delivered after execution $c=d$. 

The waiting time $w_c(h,t)$ is a delay due to resource sharing between tasks, which is a ratio of the current enqueued tasks and the available bandwidth (difference between total and utilized bandwidth):

\begin{equation}
    w_c(h,t) = \sum_{s \in D_c}\frac{\lambda_c (s) \cdot data(t)}{bw_{total}(h) - bw_{util}(h)},
\end{equation}

The communication service time $\mu_c(h,t)$ models the actual transmission between devices and remote servers. Communication is subject to the Shannon-Hartley theorem \cite{rioul2014shannon} which defines the maximum data transmitted over a noisy link. Hence, the communication service time $\mu_c(h,t)$ is:

\begin{equation} \label{eq:off-srv-time}
    \mu_c(h,t) = \frac{data(t)}{bw_{avail}(h) \cdot \log_2 (1 + \frac{p_c(h)}{n_0 \cdot bw_{avail}(h)})}
\end{equation}

where $n_0$ is the noise spectral density and $p_c$ channel transmission power. Finally, the total communication latency $T_c(h,t)$ is a sum of communication waiting time $\mu_c(h)$ and service time $w_c(h)$ such as $T_c(h,t) = \mu_c(h,t) + w_c(h,t)$.


\subsubsection{Execution latency}
The execution on the shared infrastructure is represented as a queuing M/M/1 network. Each server is a queue with independent rates and is interconnected with other queues to form a network~\cite{bramson2008stability}. Remote server $n \in N$ utilization is accumulated load and is defined as:

\begin{equation}
    U_e(v) = \sum_{s \in S}\sum_{t \in \mathcal{T}(v)} \frac{\lambda_e(s) \cdot MI(t)}{MIPS(v)},
\end{equation}

where $MI(t)$ is the number of instructions for a task $t$ and $MIPS(v)$ represents the capacity in terms of millions of instructions per second, $\lambda_e(v)$ is the arrival rate of the task and $\mathcal{T}(v)$ is a set of tasks assigned to the server $v$.

The waiting time $w_e(v)$ is the delay in task execution due to resource contention and is defined as:

\begin{equation}
    w_e(v,t) = \frac{\sum_{s \in S} \lambda_e(s) \cdot MI(t)}{1 - U(v)}
\end{equation}

The actual execution is defined as the service time $\mu_e(n,t)$, which is the ratio between the task load $t$ and the server capacity $v$:

\begin{equation}
	\mu_e(v,t) = \frac{MI(t)}{MIPS(v)},
 \end{equation}

Finally, we define execution latency $T_e(v,t)$ based on the execution waiting and service times as $T_e(v,t) = w_e(v,t) + \mu_e(v,t)$.




\subsection{Battery lifetime} \label{sec:battery-lifetime}
Mobile devices are battery-powered, thus energy saving is critical. We introduce energy models of local execution and network transmission, major drivers of mobile energy consumption\cite{tawalbeh2016studying}. We assume a mobile multicore execution power model \cite{ali2016mobile} with power states \cite{zhang2017enabling}:



\begin{equation}
    p_{e}(m) = b_{cores(m)} + \sum_{i=0}^{cores(m)} (\beta_{U_e(m)} \cdot U_e(m)) + \beta_{base} \cdot \frac{T_{idle}}{C}
    \label{eq:pcomp}
\end{equation}

where $cores(m)$ is the number of CPU cores on a mobile device $m$, $U_e(m)$ utilization per core, $\beta_{U_e(m)}$ and $\beta_{base}$ are energy coefficients for the operating and idle power states, $b_{cores(m)}$ is a CPU power baseline for a specific number of cores, $T_{idle}$ and $C$ are idle state time duration and number of power state transitions. 


The power model for network transmission $p_{c}(h_m)$ is derived from the Shannon-Hartley theorem:

\begin{equation}
    p_{c}(h_m) = n_0 \cdot bw_{avail}(h_m) \cdot (2^{\frac{Ch(h_m)}{bw_{avail}(h_m)}} - 1).
\end{equation}

where $bw_{avail}(h_m)$ is the bandwidth on the channel $h_m$ of the mobile device $m$, and $Ch(h_m)$ is a channel capacity that is an effective limit on bandwidth due to noise. Subsequently, we can define the total energy consumption on the mobile device as the sum of local execution and transmission energy consumption models, defined as $E(v,t,m,h_m) = T_e(v,t) \cdot p_{e}(m) + T_{c}(h_m) \cdot p_{c}(h_m)$. Finally, the battery lifetime of the device $BL(v,t,m,h_m)$ is defined as the ratio between the differentiation of the full battery $bcap$ and the total energy consumption until the time instant $\tau$ and full battery capacity as $BL(v,t,m,h_m) = \frac{bcap - \sum_{\tau}E(v,t,m,h_m)}{bcap}$.




\subsection{Resource utilization cost} \label{sec:pricing}
Edge and cloud are commercial services that bring monetary costs to mobile users who utilize resources owned by resource providers. Including the monetary objective in the decision-making is important to validate the approach in practical commercial environments where budgetary constraints can impact performance. The utilization cost is defined as:

\begin{equation} \label{eq:resource-pricing}
    PR(v,t) = 
    \begin{cases}
        0 & \text{if local} \\
        T_e(v,t) \cdot cost_r(v,t) & \text{if cloud} \\
        T_e(v,t) \cdot (cost_r(v,t) + cost_e) & \text{if edge} \\
    \end{cases}
\end{equation}

The first case of local execution has no cost, since no remote resources are rented. The second case brings cost when cloud resources are rented for task execution latency time $T_e(v,t)$. The cloud price $cost_r(v,t)$ for the execution task $t$ on the target server $v$ is defined:

\begin{equation}
    cost_r(v,t) = cost_{cores}(v) \cdot MI(t) + cost_{stor}(v) \cdot data(t)
\end{equation}

where $cost_{cores}(v)$, $cost_{stor}(v)$, and $data(t)$ represent cost units for CPU and data storage. The third case accounts for renting edge servers for execution latency time $T_e(v,t)$ where the price includes edge price penalty $cost_e$ for using low-latency service \cite{de2019multi}.

\section{FRESCO Offloading Solution} \label{sec:opt-problem}

\subsection{Reputation state manager} \label{sec:reputation}
The blockchain-based reputation state manager distributes task incentives to encourage servers' participation in resource-sharing and successful task completion. Task incentives are computed based on the task completion time, meaning that shorter completion times result in higher rewards. The rewards stimulate servers to perform task executions reliably and efficiently, and compete with each other by offering better performance. 
The task incentive $inc_{\tau}(v,t,h)$ at time instant $\tau$ is defined as:

\begin{equation} \label{eq:incentive}
    inc_{\tau}(v,t,h) = max\{\frac{\nabla - RT(v,t,h)}{\nabla}, 0\}
\end{equation}
where $\nabla$ is a timing constraint. 
The task incentive is normalized $[0,1]$ to prevent potential blockchain overflow.

The reputation model has to adhere to blockchain consensus restrictions. The consensus requires that on-chain updates are deterministic, to reach an agreement between blockchain nodes. Therefore, stochastic and floating-point arithmetic is not allowed on-chain \cite{battah2021blockchain}. Also, resource and time consumption on the blockchain is limited to prevent resource saturation. To address the consensus determinism requirement and limited resource consumption, we define a linear reputation model:

\begin{equation} \label{eq:reputation-short}
    R_{\tau}(v,t,h) = R_{\tau-1}(v,t,h) \cdot (1 - \omega) + \omega \cdot inc_{\tau}(v,t,h)
\end{equation}

where $R_{\tau}(v,t,h)$ is the current reputation score, $R_{\tau-1}(v,t,h)$ is a previous reputation score, and $\omega$ is a weight factor to balance between new and old reputation. Although the model stores only the last score reputation, implicitly, it accounts for multiple past values. It can be expanded to the equivalent formula, which tracks historical reputation performance by storing past reputation scores:

\begin{align}
    R_{\tau}(v,t,h)  = inc_{1}(v,t,h) \cdot (1 - \omega)^{\tau - 1} + \nonumber\\
    \sum_{i = 0}^{\tau - 2} \omega (1 - \omega)^i inc_{\tau-1}(v,t,h) 
\end{align}


Reputation scoring ensures that only reliable servers are selected for offloading. Combining both incentives and reputation scores ensures a balanced trade-off where reputation is used as a long-term performance indicator and incentives as immediate short-term rewards to stimulate continuous improvement in server reliability and prioritize reliable servers.

To summarize, the presented reputation-incentive dual approach is encoded as an HSC on the blockchain to assess the reliability level of servers based on past performance. It also ensures trust against reputation malicious tampering for gaining incentives unfairly. The reputation update is according to the presented reputation model based on provided time measurements that are acquired off-chain from mobile devices. 

\subsection{Offloading decision engine} 
\revnote{R2\_C3}{The end goal is to minimize application response time and resource utilization costs and maximize device battery by offloading tasks efficiently. Individual objectives are formally transformed into a score-based constraint optimization problem:}

\begin{equation} \label{eq:optimization}
    \begin{aligned}
&\rev{\min \sum_{t \in A} \sum_{v \in V} \; score(v, t, m, h)} \\
&\rev{\text{where } score(v, t, m, h) = 
\alpha \cdot \left( RT(v, t, h) - \widehat{RT}(v, t, h) \right)} \\
&\rev{\quad + \beta \cdot \left( \widehat{BL}(v, t, m, h_m) - BL(v, t, m, h_m) \right)} \\
&\rev{\quad + \gamma \cdot \left( PR(v, t) - \widehat{PR}(v, t) \right)} \\
&\rev{\text{s.t.} \quad RT(v, t, h) \leq \nabla, \quad \forall v \in V,\; \forall t \in A} \\
&\rev{\quad BL(v, t, m, h_m) > 0, \quad \forall v \in V,\; \forall t \in A} \\
&\rev{\quad PR(v, t) \leq p_r, \quad \forall v \in V,\; \forall t \in A} \\
&\rev{\quad AP_\tau \leq D}
    \end{aligned}
\end{equation}

\rev{where $score$ is a linear combination of the objectives. $\alpha$, $\beta$, and $\gamma$ are user-defined weights for response, battery, and resource cost respectively ($\alpha$ + $\beta$ + $\gamma$ = 1). Objectives with caret symbols are local optimum values. The goal is to find a server $n$ that minimizes the value of the $score$}. The weight factors can be fine-tuned according to user preferences and subject to sensitivity analysis. Furthermore, $A$ is a task set of certain application, and $AP_{\tau}$ is an overall application response time until $\tau$ time instant. $\nabla$, $D$ and $pr$ represent task timing constraint, application time deadline, and price constraint that can be application-dependant (e.g., 1500 ms reaction time in a traffic safety \cite{lujic2021increasing}), user-defined, or defined by developers for testing purposes. Battery lifetime is limited on mobile device $m$; thus, the goal is to avoid total discharge.

\subsubsection{SMT encoding} \label{sec:smt-modelling}
Encoding is necessary to translate Equations~\ref{eq:optimization} into a form, known as SMT formulas, that a target solver can automatically solve. The SMT combines first-order Boolean logic with constraint programming to express resource constraints and deadlines of real-time system \cite{cheng2016smt}. The SMT is lighter than machine learning solutions that are usually exposed as central third-party services \cite{lin2020survey}, and it is suitable for less powerful devices \cite{avalascai2021}. Additionally, we encode infrastructure capacities, task requirements, and servers' reputation as in Equation~\ref{eq:smt-encoding}. It combines them all and uses an SMT solver to find a reliable edge server.

\begin{equation} \label{eq:smt-encoding}
    \begin{aligned}
        reputation: (R_{\tau}(v,t,h) \geq rp) \wedge (0 \leq rp \leq 1) \\
        batteryLife: (BL(v,t,m,h_m) \cdot bcap - E(v,t,m,h_m)) \geq 0 \\
        storageLimit: \sum_{t \in O_\tau} data(t)\leq stor(v) \\
        cpuLimit: \sum_{t \in O_\tau} MI(t) \leq cpu(v) \\
        memoryLimit:  \sum_{t \in O_\tau} mem(t) \leq mem(v) \\
        taskReady: \sum_{t \in O_\tau} (\delta_{in} (t) = \emptyset \wedge t \notin O_{<\tau})\\
    \end{aligned}
\end{equation}

The \textit{reputation} constraint refers to server reputation which has to be above a certain threshold. To determine the reputation threshold $rp$, we apply $k$ criteria from \cite{iqbal2020blockchain} where top $k$ servers with the highest reputation score will be considered. We take a reputation score, which is minimum among $k$ servers as the reputation threshold $rp$. Here, the \textit{batteryLife} constraint verifies that the mobile device's battery is not drained completely. The \textit{storageLimit} constraint verifies that the input and output data of all offloaded tasks $O_{\tau}$ until time instant $\tau$ does not exceed storage capacity on the target server $v$. Similarly, CPU and memory capacities are labeled as \textit{cpuLimit} and \textit{memoryLimit} respectively. Finally, the \textit{taskReady} label indicates that the application task is ready for offloading only when tasks' input dependencies $\delta_{in}(t)$ on prior tasks are completed (i.e., empty set) and the current task $t$ was not part of a previous executed task set $O_{<\tau}$ before time instant $\tau$. Finally, we combine Equation~\ref{eq:optimization}, and~\ref{eq:smt-encoding} with logical AND operator into a single SMT logical formula. The final result of verifying the formula should be a reliable server location for offloading. However, solving the optimization function in Equation \ref{eq:optimization} is NP-hard, which is very time-consuming and impractical for real-time systems. We propose an online algorithm based on a heuristic in the next section, which can find a feasible solution in a reasonable amount of time.

\subsection{FRESCO Algorithm} \label{sec:algorithm}

\begin{algorithm}[t]
\scriptsize
    \caption{FRESCO Algorithm} \label{alg:edge_off_algo}
    \begin{algorithmic}[1]
        \Procedure{FRESCO}{$candList, currSite, reps, tasks, constr, \alpha, \beta, \gamma$}
            \State $transactions = list()$ \label{trx-declaration}
            \For{each $task$ in $tasks$} \label{for-tasks}
                \For{each $candSite$ in $candList$} \label{for-cand-list-first-for}
                    \If{$RT(task, candSite, currSite) \leq optRT$} 
                    \State $optRT = RT(task, candSite, currSite)$\label{optimum-rt}
                    \EndIf
                     \If{$EC(task, candSite, currSite) \leq optEC$} 
                    \State $optEC = EC(task, candSite, currSite)$
                    \EndIf
                     \If{$PR(task, candSite, currSite) \leq optPR$} 
                    \State $optRT = PR(task, candSite, currSite)$\label{optimum-price}
                    \EndIf
                \EndFor
                \For{each $candSite$ in $candList$} \label{for-cand-list-second-for}
                    \State \parbox[t]{200pt}{$score(candSite) = \alpha(RT(task,candSite,currSite) - optRT) + \beta(EC(task,candSite,currSite) - optEC) + \gamma(PR(task,candSite,currSite)-optPR)$\strut} \label{score}
                \EndFor
                \Do \label{do-while-loop}
                    \If{$candList.empty()$} \label{empty-cand-list}
                        \State \textbf{break}
                    \EndIf
                    \State $selSite = SMTSOLVING(score,candList, reps,constr)$ \label{smt-solving}
                    \If{$OFFLOAD(selSite, task)$}
                        \State $d = compTaskConstrMeasure(selSite, task)$ \label{task-deadline}
                        \State $transactions.append((d, selSite))$ \label{trx-append}
                        \State \textbf{break}
                    \EndIf
                    \State $transactions.append((0, selSite))$\label{null-incentive}
                    \State $score.pop(selSite)$
                    \State $candList.pop(selSite)$
                    \State $reps.pop(selSite)$ \label{pop-reputation}
                \doWhile {True}
            \EndFor\\
            \Return $transactions$ \label{return-trx}
        \EndProcedure
    \end{algorithmic}
\end{algorithm}

The offloading algorithm needs to solve the objective function and respect application deadlines, task timing constraints, and resource limitations. Therefore, we propose the FRESCO algorithm (Algorithm~\ref{alg:edge_off_algo}) for performing reliable edge offloading decisions. FRESCO algorithm represents an offloading decision engine executed on a mobile device. The design decision is deliberate to maintain mobile device autonomy without relying on unreliable and failure-prone edge servers. Reputation scores are obtained from the HSC-based reputation state manager, which does not need consensus for reputation retrieval but only for reputation update after task offloading execution. Inputs for FRESCO algorithms are the list of candidate servers, the server where the previous task was executed ($currSite$), the reputation scores per server $reps$, a list of tasks $tasks$, a list of constraints $constr$, and user-defined weights $\alpha$, $\beta$ and $\gamma$. First, we declare a transaction list recording every offloading attempt and its associated constraint (line~\ref{trx-declaration}). Then, we compute local optima for each objective (lines~\ref{optimum-rt}-\ref{optimum-price}), which are used to calculate servers' optimization score (line~\ref{score}) (first \textit{for} loop on line~\ref{for-tasks}). We iterate until the candidate list is empty or the task is successfully offloaded (\textit{do-while} loop on line~\ref{do-while-loop}). If the candidate list is empty (line~\ref{empty-cand-list}) then it exits from the \textit{do-while} loop and returns accumulated transactions. Otherwise, the SMT solver on line~\ref{smt-solving} selects the server. If offloading fails, then the server is removed from the candidate list and its associate objective values (lines~\ref{null-incentive}-\ref{pop-reputation}) and loops back on line~\ref{do-while-loop}. If offloading succeeds, the difference between execution time and the constraint $\nabla$ is computed (line~\ref{task-deadline}) and appended to the transaction list (line~\ref{trx-append}). The transaction list is returned (line~\ref{return-trx}) to the HSC for reputation update.

The computational complexity of FRESCO depends on $|T|$, which is the cardinality of the set of application tasks $T$, and $|N|$, which is the cardinality of the set of nodes $N$. This can be seen by the \textit{for} loop on line~\ref{for-tasks}, that is executed $|T|$ times, and \textit{for} loops on lines~\ref{for-cand-list-first-for},~\ref{for-cand-list-second-for}, that iterate over $N$ set. Also, \textit{do-while} loop on line~\ref{do-while-loop} is executed $|N|$ times in the worst case. \revnote{R2\_C4}{However, the most impacting factor on FRESCO complexity is the complexity of the SMT solver (SMTSOLVING function on line~\ref{smt-solving}). SMT solving generalizes the Boolean satisfiability problem (SAT), which makes it NP-hard. General exact computational complexity of SMT solving is hard to deduce since it depends on multiple factors, such as heuristic space search, clause learning, problem size, and structure~\cite{smttheory}. Therefore, many researchers turn to empirical benchmarking, where the selection of an SMT solver engine has a strong impact on performance~\cite{hofler2014smt}. Despite SMT solver being used in practice for real-time task scheduling on embedded and mobile devices~\cite{avalascai2021,cheng2016smt}, we are compelled to empirically validate SMT solver's fitness for usage on resource-constrained edge devices for latency-sensitive mobile applications. Lastly, FRESCO computational complexity is independent of the HSC-based reputation state manager, which is executed on blockchain and has a linear relationship with $transactions$, which is a tuple list of response time and associated server (line~\ref{return-trx}). When the tuple list is computed, it is forwarded to the HSC-based reputation state manager for reputation update on the public blockchain}. 


\section{Experimental Evaluation} \label{sec:experiment}
\subsection{Implementation and testbed}
Python v3.8 simulator of edge-cloud infrastructure, Z3 SMT-based offloading decision engine, Ganache \footnote{https://archive.trufflesuite.com/ganache/} blockchain emulator with real-world Solidity v0.9.0\footnote{https://soliditylang.org/} HSC-based reputation state manager are deployed on an AMD64 server with a 40-core 1.80GHz CPU and 128Gb RAM. 
The decision engine connects to the Ganache when reputation needs to be updated and stored during runtime, and offloads DAG-based application tasks on simulated edge-cloud clusters during mobility between different cells. The infrastructure is simulated based on the OpenCellID dataset \cite{opencellid}, which contains cell tower locations distributed over vast areas. The workload on the nodes is simulated through the queuing network (Section~\ref{sec:edge-queue-model}). 
Using an emulator instead of a real blockchain is due to the limited number of Ethereum tokens available, which prevents repeated experiments for statistical significance. We assume Proof-of-Authority (PoA) consensus, popular in both private and public Ethereum whose consensus delay is around 4 seconds \cite{ethereumblocktime}. Developers usually use this type of consensus to get easy access and fast feedback. We have open-sourced our prototype publicly\footnote{\url{https://github.com/jzilic1991/hybrid-edge-blockchain}}. 

\subsection{Experimental design and setup}
\subsubsection{Computing and networking infrastructure} \label{sec:comp-net-specs}
Table~\ref{tab:comp-specs} shows the target infrastructure configuration. It reflects our infrastructure's configurations of different edge, cloud, and mobile devices. We classified servers into several classes to capture resource heterogeneity. The mobile device has limited resources compared to other nodes. The ED is an edge database server that has fast-speed network access and large data storage capacity to handle data-intensive (DI) tasks; the second one represents a computational-intensive server (EC) that has a high number of CPU cores to cope with computational-intensive (CI) tasks, and the third one represents an edge regular server (ER) with moderate resource capacities. The cloud server is the most resourceful one, but has higher latency.

\begin{table}[t]
    \caption{Computing infrastructure} \label{tab:comp-specs}
    \begin{center}
        \begin{tabular}{ | m{2cm} | m{1cm}| m{1cm} | m{1cm} | m{1cm} |} 
            \hline
            \textbf{Node class} & \textbf{CPU cores} & \textbf{CPU (GHz)} & \textbf{RAM (GB)} & \textbf{Storage (GB)} \\ \hline
            ED server & 8 & 2100 & 8 & 300\\
            \hline
            EC server & 16 & 2800 & 16 & 150\\ 
            \hline
            ER server & 4 & 1800 & 8 & 150\\
            \hline
            CD server & 64 & 2400 & 128 & 1000\\
            \hline
            Mobile device & 2 & 1800 & 8 & 16\\
            \hline
        \end{tabular}
    \end{center}
\end{table}

\begin{table}[t]
  \caption{Empirical latency measurements as constraints and deadlines from real-world applications in milliseconds}
  \label{tab:proc-lat-measurements}
  \begin{tabular}{|l|l|l|l|l|l|l|}
    \hline
    \multirow{2}{*}{} &
      \multicolumn{2}{c|}{\textbf{Intra($D$=108)}} &
      \multicolumn{2}{c|}{\textbf{MobiAR($D$=400)}} &
      \multicolumn{2}{c|}{\textbf{NaviAR($D$=800)}} \\
    \hline
    $\nabla$ & Proc & Net & Proc & Net & Proc & Net\\
    \hline
    Edge & 18 & 15 & 2-20 & 15 & 250-300 & 300-400 \\
    \hline
    Cloud & 2-20 & 90 & 1 & 300 & 2-20 & 1000-1500 \\
    \hline
    Mobile & 300 & 0 & 300 & 0 & 800 & 0 \\
    \hline
  \end{tabular}
\end{table}

We adopt processing and network latencies as application QoS deadlines from three real-world use cases, described in Table~\ref{tab:proc-lat-measurements}. In Table~\ref{tab:proc-lat-measurements}, "Proc" indicates the processing timing constraint, while "Net" is the networking timing constraint. Note that we distinguish task timing constraint $\nabla$ from application deadline $D$. We measure QoS violations against application deadlines. 

\subsubsection{Mobile DAG applications}
Mobile applications are modeled as DAGs, which is a common method of mobile application modeling ~\cite{avalascai2021,zilic2022edge}. 
These applications exhibit a pipeline workflow structure, which is typical for AI-based applications. Table~\ref{tab:task-specs} specifies task categories from which the applications are constructed, while Tables~\ref{tab:intrasafed-specs},~\ref{tab:MobiAR-specs}, and~\ref{tab:NaviAR-specs} describe structures of selected applications. We selected the following applications because they are latency-sensitive and are part of an emerging market where edge computing is a key technology enabler. 

\textbf{(i) Intrasafed:} It is a traffic safety application~\cite{lujic2021increasing}, which employs an AI-based object detection that detects pedestrians at intersections, notifying drivers in real-time to prevent accidents. We simulated the application in our simulator with latency measurements from the original work, presented in Table~\ref{tab:proc-lat-measurements}. It has a deadline of $D$ = 108 ms for the average driver's notification latency via 5G networks. 
\textbf{(ii) MobiAR:} It is a generic AR object detection application \cite{ren2021adaptive}, which we extracted its application structure and executed in our simulator. The real latency measurements are extracted from the work and presented in Table~\ref{tab:proc-lat-measurements}. The application requires a deadline of $D$ = 400 ms to meet the applications' inference latency.
\textbf{(iii) NaviAR:} It is an AR live navigation executed on AR HoloLens glasses~\cite{wang2021context}. We simulated the structure in our simulator, backed by latency measurements as constraints listed in Table~\ref{tab:proc-lat-measurements}. It requires a deadline of 800 ms, which is equal to the local execution time on AR glasses.

\begin{table}[t]
\centering
    \caption{Task specifications} \label{tab:task-specs}
    \begin{tabular}{ | m{1.1cm} | m{2.1cm}| m{1.5cm} | m{1.7cm} | } 
        \hline
        \textbf{Type} & \textbf{CPU} & \textbf{Input data} & \textbf{Output data} \\ 
        \hline
        DI & 100-200 M cycles & 15-20 KB & 25-30 KB \\
        \hline
        CI & 550-650 M cycles & 4-8 KB & 4-8 KB \\ 
        \hline
        Moderate & 100-200 M cycles & 4-8 KB & 4-8 KB \\
        \hline
    \end{tabular}
\end{table}

\begin{table}[t]
\centering
    \caption{Intrasafed task specifications} \label{tab:intrasafed-specs}
    \begin{tabular}{ | m{2.8cm} | m{1.3cm}| m{0.8cm} | m{1.6cm} | } 
        \hline
        \textbf{Task} & \textbf{Type} & \textbf{RAM} & \textbf{Offloadable}\\
        \hline
        LOAD\_MODEL & Moderate & 1 GB & False\\
        \hline
        UPLOAD & DI & 1 GB & True\\
        \hline
        ANALYZE & CI & 4 GB & True\\
        \hline
        AGGREGATE & CI & 2 GB & True\\
        \hline
        SEND\_ALERT & Moderate & 1 GB & True\\
        \hline
    \end{tabular}
\end{table}

\begin{table}[t]
\centering
    \caption{MobiAR task specifications} \label{tab:MobiAR-specs}
    \begin{tabular}{ | m{2.8cm} | m{1.3cm}| m{0.8cm} | m{1.6cm} | } 
        \hline
        \textbf{Task} & \textbf{Type} & \textbf{RAM} & \textbf{Offloadable}\\
        \hline
        UPLOAD & Moderate & 1 GB & False\\
        \hline
        EXTRACT & CI & 2 GB & True\\
        \hline
        PROCESS & CI & 2 GB & True\\
        \hline
        DATA & DI & 1 GB & True\\
        \hline
        DOWNLOAD & DI & 1 GB & False\\
        \hline
    \end{tabular}
\end{table}

\begin{table}[t]
\centering
    \caption{NaviAR task specifications} \label{tab:NaviAR-specs}
    \begin{tabular}{ | m{2.8cm} | m{1.3cm}| m{0.8cm} | m{1.6cm} | } 
        \hline
        \textbf{Task} & \textbf{Type} & \textbf{RAM} & \textbf{Offloadable}\\
        \hline
        MAP & DI & 1 GB & True\\
        \hline
        GUI & Moderate & 1 GB & False\\
        \hline
        COORDINATION & CI & 4 GB & True\\
        \hline
        SHORTEST\_PATH & CI & 2 GB & True\\
        \hline
        MOTION\_COMMAND & CI & 1 GB & True\\
        \hline
        VIRTUAL\_GUIDANCE & Moderate & 1 GB & False\\
        \hline
        RUNTIME\_LOCATION & CI & 1 GB & True\\
        \hline
        DISPLAY & Moderate & 1 GB & False\\
        \hline
    \end{tabular}
\end{table}

\begin{table}[t]
    \caption{Simulation and algorithmic parameters} \label{tab:coeff}
    \centering
    \begin{tabular}{ | m{1.5cm} | m{1.5cm} | } 
        \hline
        \textbf{Parameter} & \textbf{Value}\\
        \hline
        $\lambda$ & [10, 20] \\
        \hline
        $\alpha$ & 0.5\\
        \hline
        $\beta$ & 0.4\\
        \hline
        $\gamma$ & 0.1 \\
        \hline
        $BL$ & 1000\\
        \hline
        $\omega$ & 0.3\\
        \hline
        $cost_{cores}$ & 0.023 \\
        \hline
        $cost_{stor}$ & 0.776\\
        \hline
        $\beta_{base}$ & 625.25 $10^{-3}$\\
        \hline
        $\beta_{U_e}$ & 6.9305 $10^{-3}$\\
        \hline
        $p_{cores}$ & 0.073 $10^{-3}$\\
        \hline
    \end{tabular}
\end{table}

\subsubsection{Parameters} \label{sec:parameters}
Parameters used in our experiment are defined in Table~\ref{tab:coeff}. The Poisson task arrival rate $\lambda$ range is selected so it can scale to different workload intensities. $\alpha$, $\beta$, and $\gamma$ values are selected as a representative case of the user's preferences about preferring fast response and willingness to pay a higher price for it ($\alpha > \beta > \gamma$) since we target latency-sensitive applications. $BL$ is the initial device battery capacity. Reputation weight factor $\omega$ is taken from \cite{battah2021blockchain}, which accounts for a relatively conservative reputation system to mitigate volatility in a crowdsourced system. $cost$ coefficients for CPU and storage are taken from Google Cloud \cite{gcloudprice}. \revnote{R1\_C1}{0.023 price for CPU cores is interpreted as \euro{}0.000023 for virtual CPU per second associated with the compute-intensive VM instance C2-highcpu-16 in the \textit{europe-central2} region. Storage price \euro{}0.776 is reconstructed from 0.17 for persistent SSD storage, 0.22 high-availability premium for edge caching and redundancy, and 0.15 for frequent object-level operations (e.g. Class A/B GCS ops), 0.10 for data egress within the data center, and 0.075 for monitoring and logging\cite{gcloudprice}}. Energy coefficients of $\beta_{base}$, $\beta_{U_e}$ and $p_{cores}$ are taken from \cite{ali2016mobile,zhang2017enabling} which are validated against real mobile equipment. \revnote{R1\_C1}{Recent works \cite {Ohk2022PhaseBasedLP,Caiazza2023MeasuringTE} provided more realistic power estimation for cellular communication and edge computational resources. However, these energy models are device-specific and require hardware power profiling, and hence, they are not often accessible. Our edge simulations rather adopt lightweight energy models as a balance between simplicity for simulation scalability, task generalizability, and realism that is validated with real physical measurements, as noted in \cite{ali2016mobile,zhang2017enabling}}.

\subsubsection{Datasets}
We employed the Skype availability dataset\cite{guha2005experimental} to model the system's availability. The motivation for selecting the Skype dataset is that it shares edge characteristics like geo-distribution, heterogeneity, a large number of nodes, and it constitutes the middle ground in availability ratio (60-70\%) and latency (up to $\sim$50 ms) compared to other infrastructure\cite{aral2020learning}. Traces are collected over 2,081 servers for 400 days and contain availability time intervals that are associated with each node. Nodes have different lifespans, and hence they are normalized within the $[0,1]$ time range interval.  Adopting such availability datasets from distributed systems that share similar characteristics is common in edge computing research\cite{aral2020learning,samani2023proactive} due to the lack of publicly available datasets. 

\begin{figure}[h]
    \centering    
    \includegraphics[width=\columnwidth]{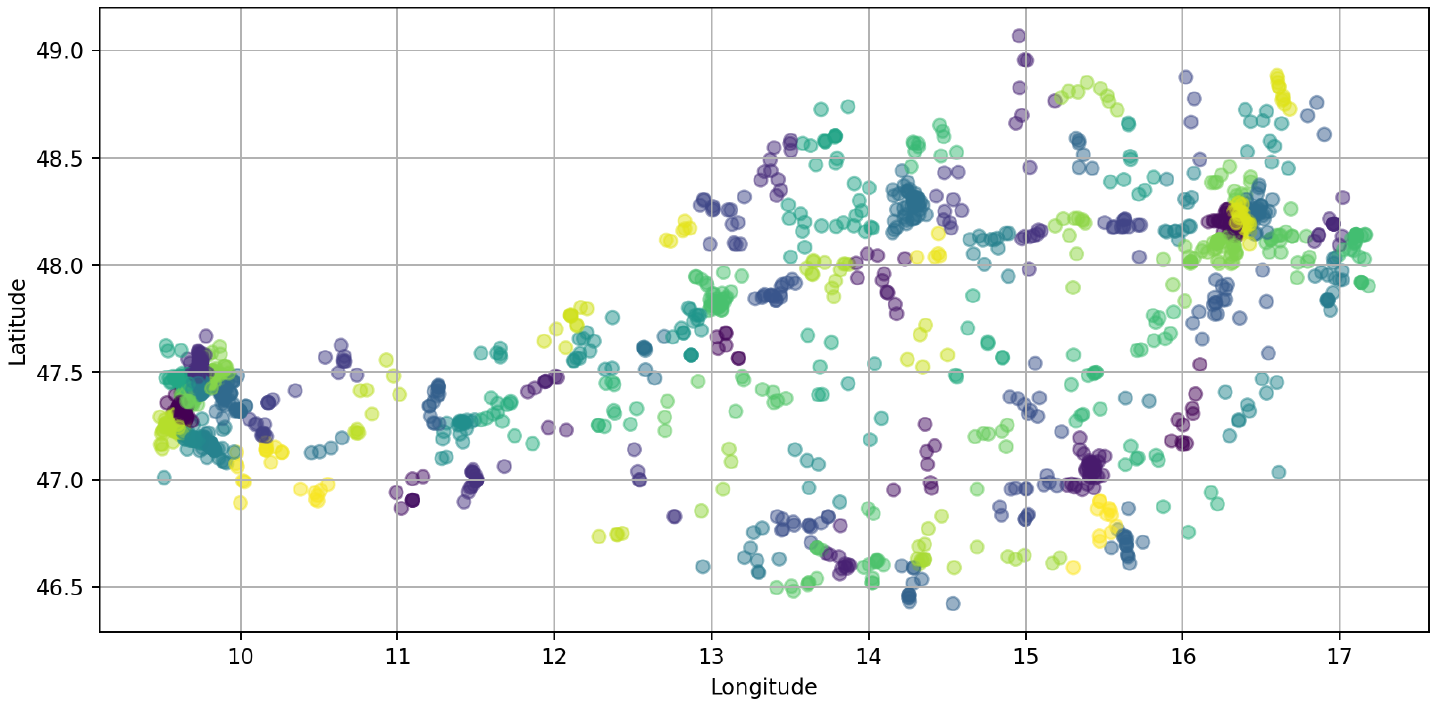}
    \caption{Cell tower locations from OpenCellID dataset ~\cite{opencellid}}
    \label{fig:at-cells}
\end{figure}

Edge and cloud deployment follow cellular base station locations from OpenCellID. OpenCellID is an open cellular database containing datasets of cell tower geolocations that mobile operators publicly publish. It is used in generating infrastructure topologies under edge computing settings \cite{topology2021mec}. We selected a dataset that contains around $3,500$ cell tower locations and randomly filtered them out to match the number of $2,081$ Skype nodes for one-to-one availability trace mapping. We clustered the entire network into $30$ cell clusters using the k-means clustering algorithm, as illustrated in Figure~\ref{fig:at-cells}. In such a deployment, location-based mobility is simulated where a mobile device visits each cell cluster and offloads tasks on remote servers. Mobile device dwelling time in each cell is evenly distributed throughout the entire simulation. Each cell cluster location has edge node classes such as ER, ED, and EC which are randomly associated with OpenCellID nodes and a single accessible cloud server. Remote servers have an associated reputation score, which is stored on a public blockchain that is globally accessible.

\revnote{R1\_C1}{We also employ LiveLab application usage traces \cite{shepard2010livelab} and AR telemetry dataset \cite{vr_ar_cg_telemetry}. We selected Livelab because it includes application frequency usage on smartphones collected over long time periods in a real-world environment, such as a university campus. To map our DAG applications with the traces, we mapped them into DAGs based on shared functionalities (e.g., vision, user interactions, event-based notifications). AR and games are mapped to MobiAR, while safety and utilities are mapped to IntraSafed, and maps and travelling are mapped to NaviAR application usage. We computed their statistical distribution where 76.3\% is mapped to MobiAR, 20\% Intrasafed, and 3.7\% is NaviAR, and use these distributions for sampling applications into \textit{random} workload. The AR telemetry dataset contains 5000 samples of uplink and downlink data during AR gaming application execution on the local computer, emulating an edge server. The dataset includes data like interarrival packet and interarrival frame interval, ideal for deriving task arrival rates and task sizes for load generation in our queueing network system. The data set is selected because two out of three DAG applications in our experiment are AR workflows, and the third application, Intrasafed, shares some similar characteristics (e.g., real-time event notifications and video streaming). Queueing task arrival rates are from $[60, 70]$ tasks, while task size is from $[10, 20]$ data. These queueing parameters, together with the aforementioned LiveLab statistical distribution, are combined into \textit{random} workload. The workload is annotated as RANDOM in the evaluation plots and separated from the parameters in Table~\ref{tab:coeff}. These real-world traces strengthens our evaluation rigor in our experiments and validate our results}.

\subsubsection{Baselines}
We compare \textit{FRESCO} with the following baselines. 
\begin{itemize}
    \item \textit{\textbf{MINLP}} is a mixed integer non-linear programming-based method that formulates constraint offloading optimization problems without reputation. The MINLP approach is the most common modeling method for offloading optimization~\cite{feng2022computation}.
    \item \revnote{R1\_C2}{\textbf{\textit{SQ}}~\cite{iqbal2020blockchain} is a blockchain-based method that considers reputation and queuing time on edge nodes in vehicular ad hoc networks, where only local and edge decisions apply.}
    \item \rev{\textit{\textbf{QRL}} is a common method based on the Markov Decision Process (MDP) model for modeling offloading \cite{lin2020survey}. Reputation is encoded as a transition probability, remote servers represent states, and objectives are modeled as reward functions. The modeling is similar to existing work that targets reliable offloading\cite{zilic2022edge}}.
\end{itemize}

\begin{figure*}[t]
    \centering
    \begin{minipage}[b]{0.32\textwidth}\centering
        \includegraphics[height=12em]{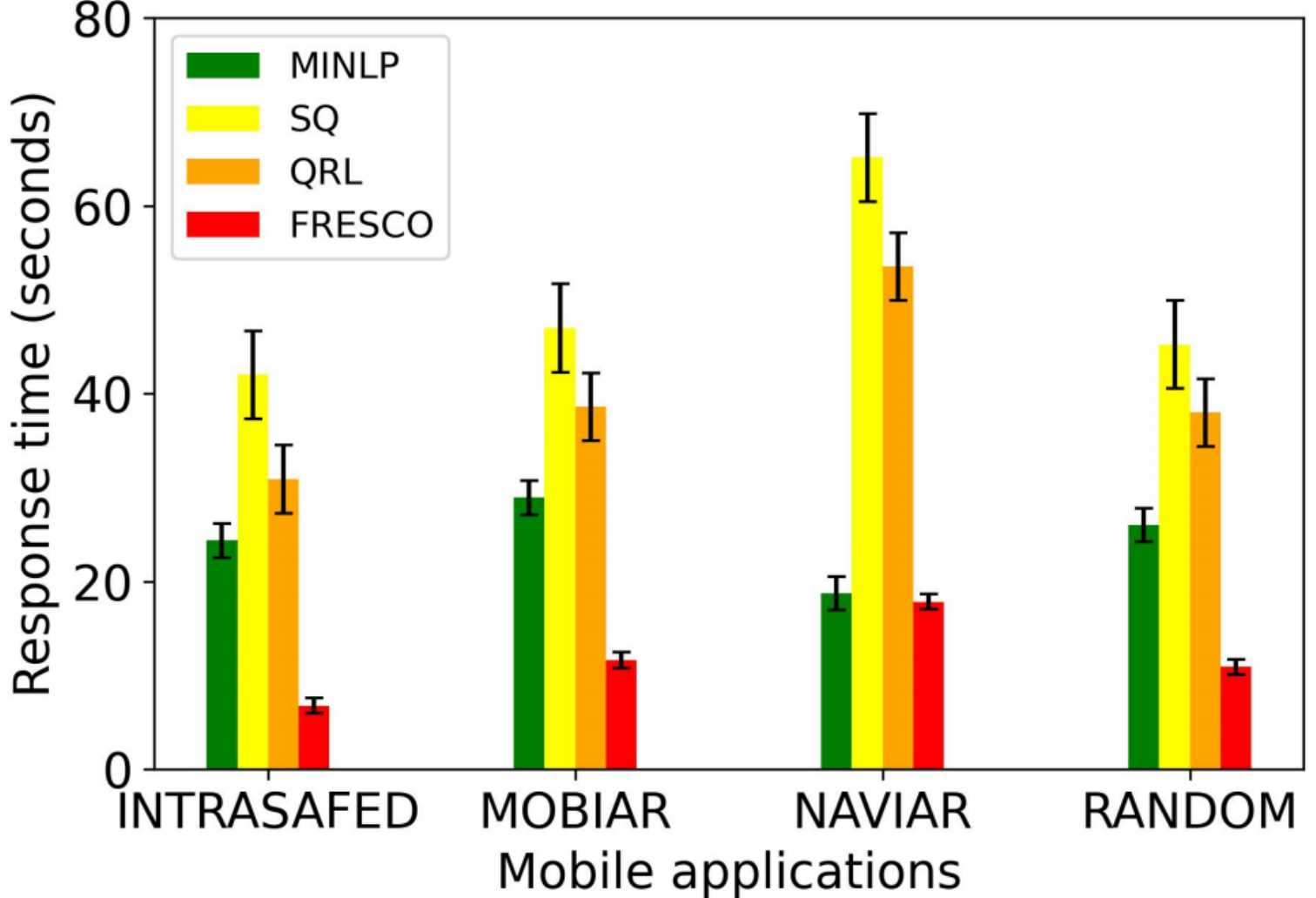}
        \caption{Response time}
        \label{fig:response-time}
    \end{minipage}
    \hfill
    \begin{minipage}[b]{0.32\textwidth}\centering
        \includegraphics[height=12em]{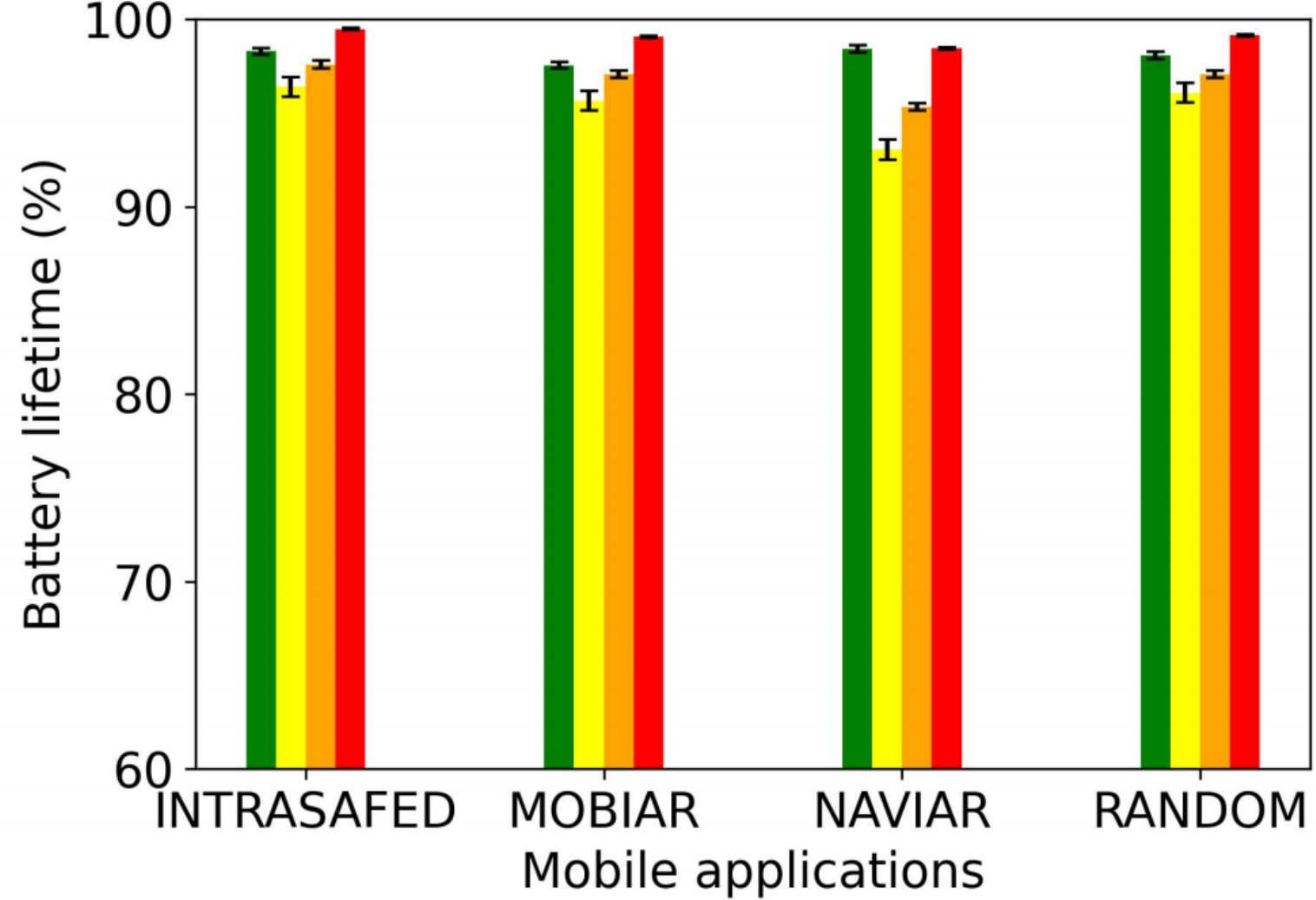}
        \caption{Battery lifetime}
        \label{fig:battery-lifetime}
    \end{minipage}
    \hfill
    \begin{minipage}[b]{0.32\textwidth}\centering
        \includegraphics[height=12em]{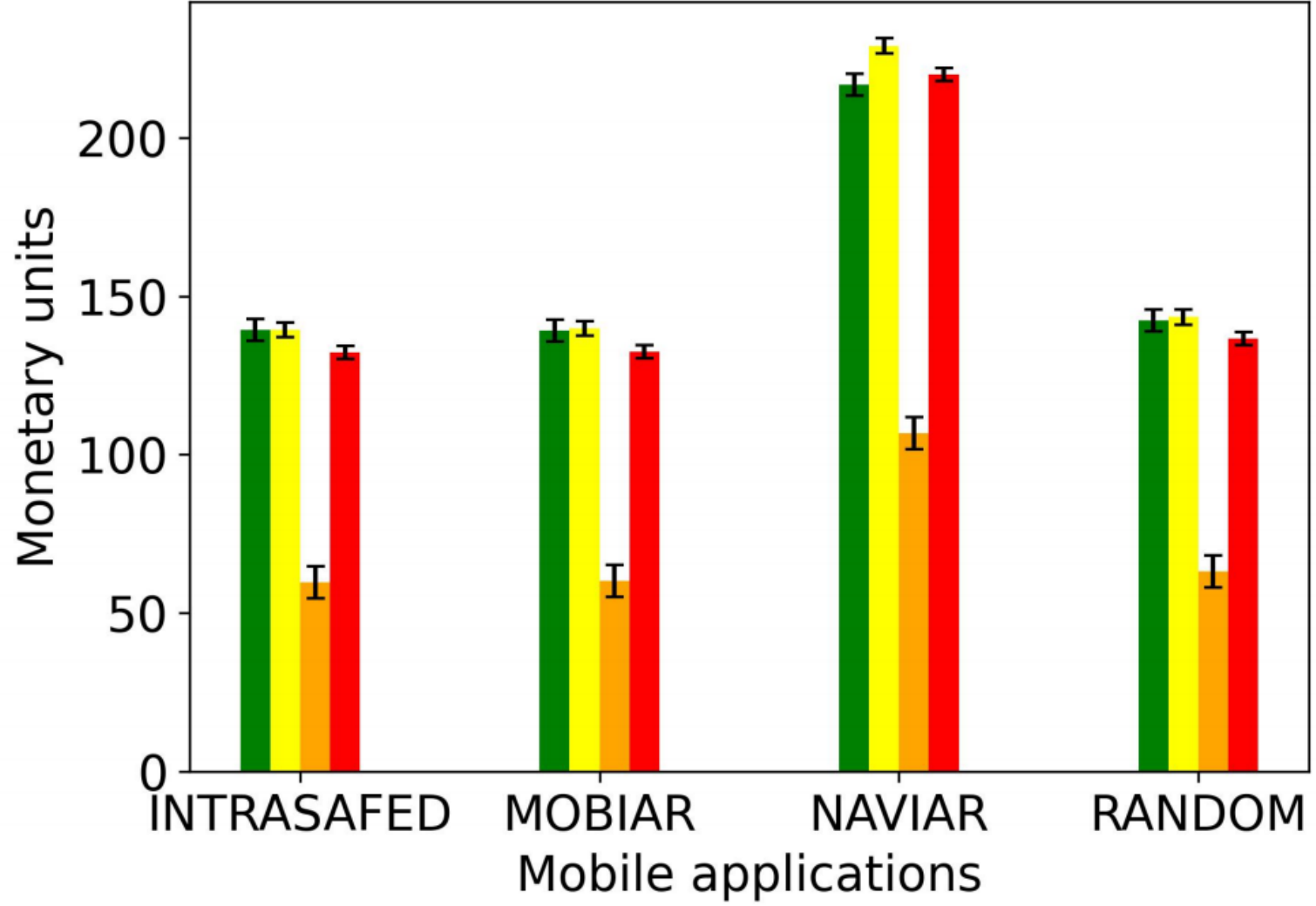}
        \caption{Resource utilization cost}
        \label{fig:resource-pricing}
    \end{minipage}
\end{figure*}

\subsection{Analysis of results}
For each experimental run, we execute 100 applications sequentially and average results over 100 runs to obtain statistically significant results.

\begin{figure*}
     \centering
     \begin{subfigure}[b]{0.3\textwidth}
         \centering
         \includegraphics[width=\textwidth]{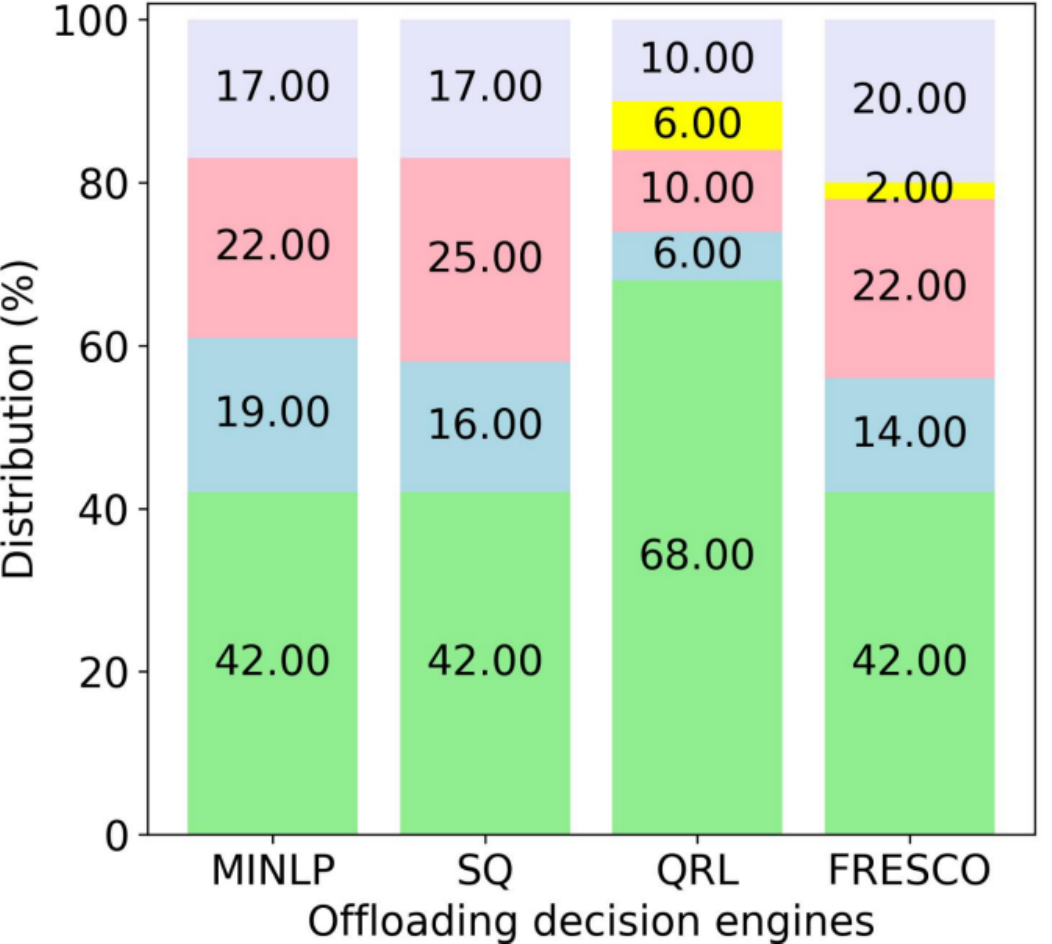}
         \caption{Intrasafed}
         \label{fig:off-dist-intra}
     \end{subfigure}
     \begin{subfigure}[b]{0.3\textwidth}
         \centering
         \includegraphics[width=\textwidth]{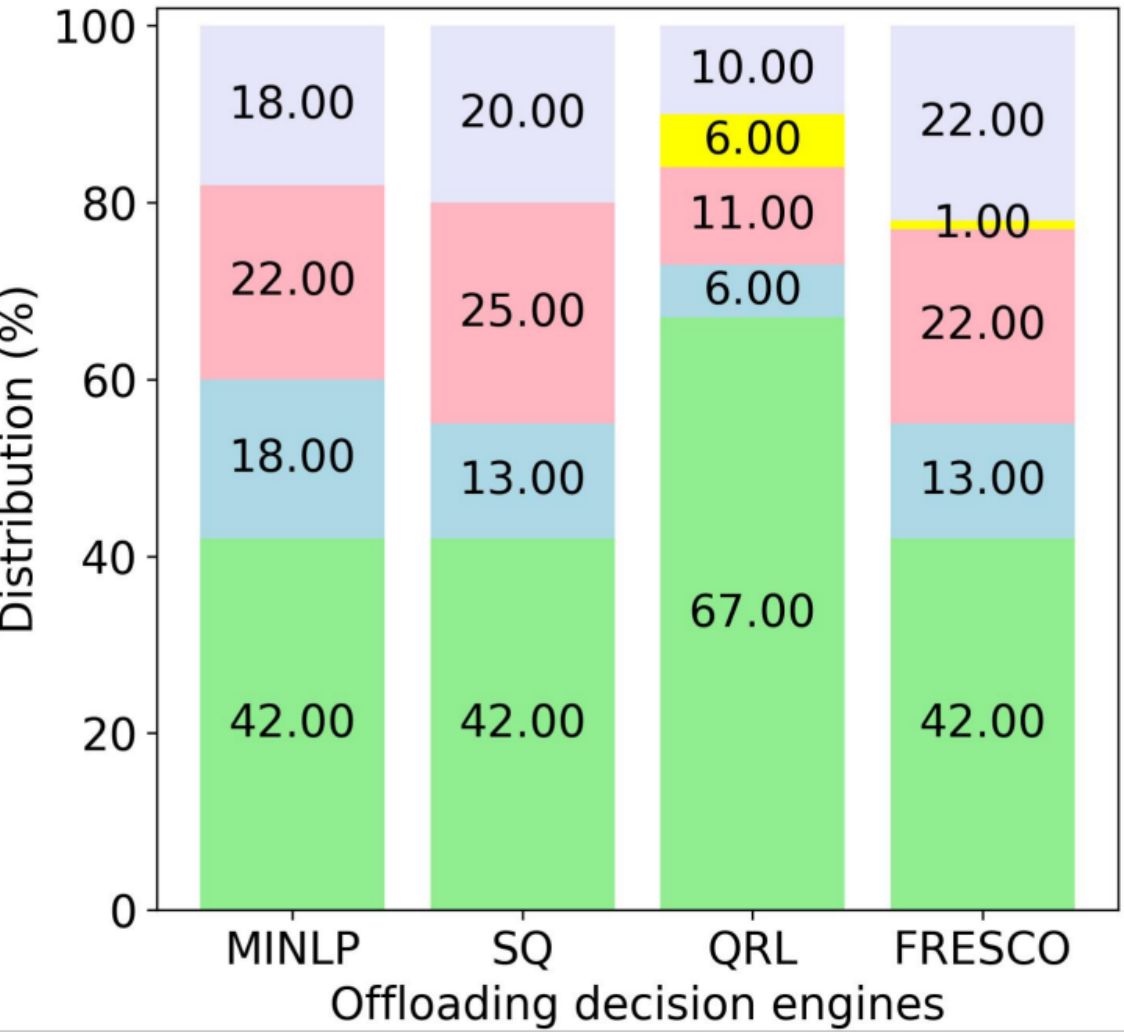}
         \caption{MobiAR}
         \label{fig:off-dist-MobiAR}
     \end{subfigure}
     \begin{subfigure}[b]{0.35\textwidth}
         \centering
         \includegraphics[width=\textwidth, height=14.5em]{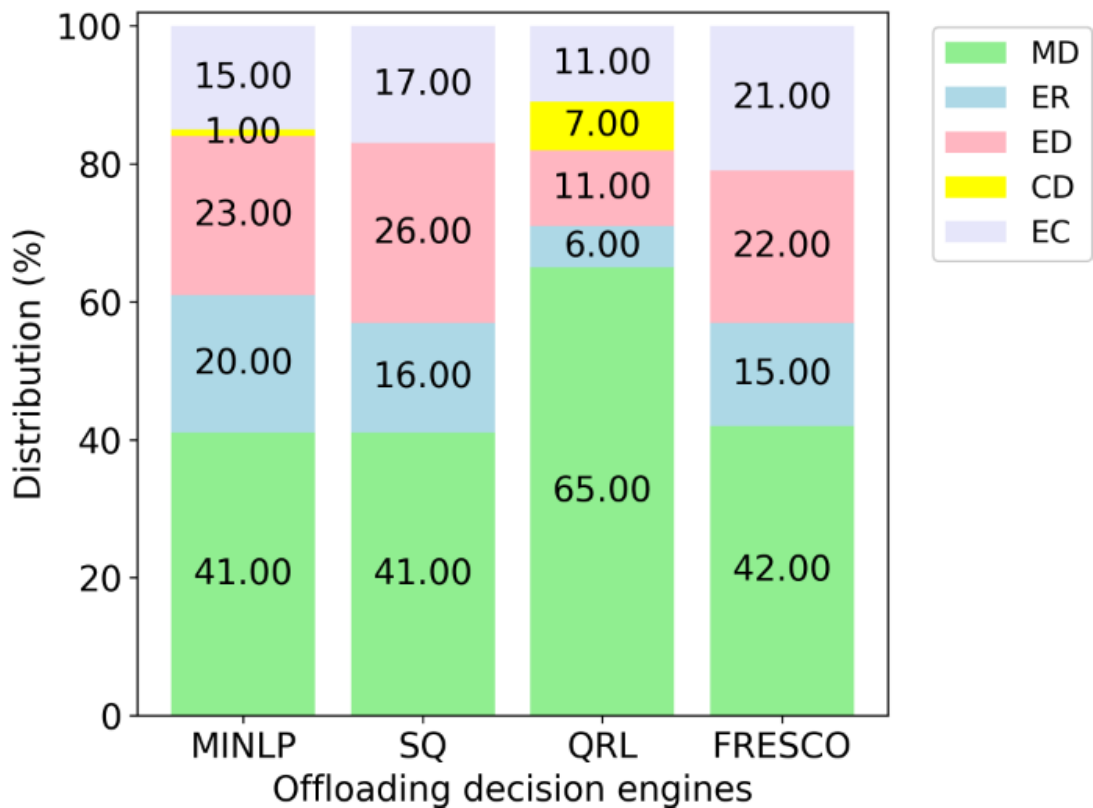}
         \caption{NaviAR}
         \label{fig:off-dist-NaviAR}
     \end{subfigure}
     \caption{Offloading distribution}
\end{figure*}

\subsubsection{\textbf{Response time}}
Figure~\ref{fig:response-time} illustrates the response time performance of offloading decision engines in Intrasafed, MobiAR, and NaviAR applications, respectively. The worst-performing decision engine is SQ, which has average response times of $41.99$, $46.96$, $65.12$, and $45.198$ seconds. SQ has a higher deviation compared to others ($6.16$, $4.03$, $5.99$  and $4.688$ seconds) which makes it more volatile, not appropriate for high reliability applications that require consistent performance. Although the SQ approach is reputation-aware and blockchain-based, its primary target is to identify malicious servers instead of reliable offloading in terms of QoS violations caused by failed or high-loaded sites. The MINLP decision engine yielded the second-best approach with $24.34$, $28.91$, $18.72$, and $26.004$ seconds due to constraint satisfaction logic inherited from the SMT solver, but reputation-oblivious. The best performance was achieved with the NaviAR application ($18.72$ seconds), which is unexpected since NaviAR has the most complex structure. The possible explanation is that the edge servers in the last visited cells were more loaded, which limits resource capacity. It could deter MINLP from taking offloading decisions on the edge and rather opt for local execution or select a far-distant cloud. $75\%$ of the offloading attempts in the last cell were concentrated on cloud and mobile. Although MINLP does not perceive reputation, selecting both mobile devices and the cloud is safe for offloading and avoids offloading failures on the failure-prone edge, which in the last two cells has limited availability ($12-25\%$). Offloading failure would impose a longer response time, as seen in Intrasafed and MobiAR applications. Lastly, our FRESCO solution outperforms other decision engines due to frequent offloading on more reliable servers, which resulted in response time performance of $6.75$, $11.61$, $17.81$, and $10.909$ seconds, and the lowest standard deviations between $0.438$ and $0.866$ seconds, which makes it also the most consistent solution.

\subsubsection{\textbf{Battery lifetime}}
Figure~\ref{fig:battery-lifetime} illustrates the battery performance of offloading decision engines in all four application cases. The SQ decision engine drains the device battery the most, with $96.38\%$, $95.33\%$, $93.34\%$, and $96.071\%$. \revnote{R1\_C2}{A higher rate of failed offloading attempts drains the energy more than the longer response time (Figure~\ref{fig:response-time}) as in QRL, whose battery lifetimes are $97.582\%$, $97.063\%$, $93.03\%$ and $97.07\%$}. MINLP and FRESCO, on the other hand, have battery lifetimes that reflect response time performance from Figure~\ref{fig:response-time}. MINLP battery lifetimes are $98.28\%$, $97.54\%$, $98.42\%$ and $98.076$, while FRESCO has the highest battery lifetime of $99.47\%$, $99.06\%$, $98.43\%$ and $99.144\%$.

\subsubsection{\textbf{Resource utilization cost}}
\revnote{R1\_C2}{Figure~\ref{fig:resource-pricing} shows the resource utilization cost of all four approaches. Here. 
QRL incurs lower resource utilization costs (ranging from $59.743$ to $106.763$ monetary units), although it has poor response time (see Figure~\ref{fig:response-time}). This is because it offloads to a highly available cloud node, which has a lower utilization cost and a cost-free mobile.} SQ is the most expensive solution due to failed offloading attempts and a fixed \textit{k} parameter, which does not scale with several nodes, and thus limits alternative servers for offloading consideration and can lead to potentially higher costs. According to SQ, best \textit{k} sites are the most reputable ones but can be highly loaded and limit re-offloading alternatives in case of failed or untimely execution. SQ monetary costs are $139.3$, $139.16$, $228.23$, and $143.409$ units for three applications. When comparing MINLP and FRESCO, FRESCO emerged as the second-best cost-effective solution and cheaper than MINLP in Intrasafed ($132.22$ vs $139.33$ units) and MobiAR ($132.38$ vs $139.08$ units) use cases, but worse when offloading NaviAR ($219.89$ vs $216.76$ units). In NaviAR's case, MINLP is slightly cheaper than FRESCO because it offloads a minor portion of tasks to the cloud, which is cheaper than edge. FRESCO did not yield the overall best cost-effectiveness because hyperparameters are tuned, so it prefers faster and energy-efficient solutions rather than low-cost sites.

\subsubsection{\textbf{Offloading distribution}}
Offloading distribution Figure~\ref{fig:off-dist-intra} for all four application cases shows the distribution of offloaded tasks to different node classes. \revnote{R1\_C2}{In the Intrasafed use case, the QRL was worse-performant because most of the offloading decisions were addressed to mobile ($68\%$) instead of expensive edge servers and thus is the cheapest solution (Figure~\ref{fig:resource-pricing}) and not necessarily the least energy-efficient (Figure~\ref{fig:battery-lifetime}).} SQ only considers \textit{k} sites with the highest reputation and selects the shortest queue waiting time. \textit{k} parameter is fixed and does not scale with several nodes, which can limit the number of alternative servers and exclude viable ones that are less loaded and sufficiently reliable. MINLP, on the other hand, has also similar offloading distribution but does not restrict its offloading options. FRESCO, comparably to the previous two aforementioned baselines, is more flexible and utilizes all site types, including cloud ($2\%$) for CI tasks when edge servers are less reliable, also less reliant on moderate ER sites ($14\%$ compared to $19\%$ and $16\%$ in MINLP and SQ respectively), and utilizes resource-rich EC sites more frequently ($20\%$ compared to $17\%$ in MINLP and SQ). FRESCO's offloading distribution composition is similar in the MobiAR case (Figure~\ref{fig:off-dist-MobiAR}) and reflects FRESCO's higher performance in both applications. In the NaviAR case (Figure~\ref{fig:off-dist-NaviAR}), MINLP and FRESCO have different offloading distribution compositions, but performance-wise are comparable (Figure~\ref{fig:response-time}). FRESCO balances reliability and performance, where the most reputable servers are not necessarily efficient ones. MINLP is reputation-oblivious, but selecting the most efficient servers can be beneficial sometimes if the underlying infrastructure is more reliable and has fewer failures or less volatile load.

\subsubsection{\textbf{QoS violations}}
\begin{figure}[h]
    \centering
    \includegraphics[height=13.5em]{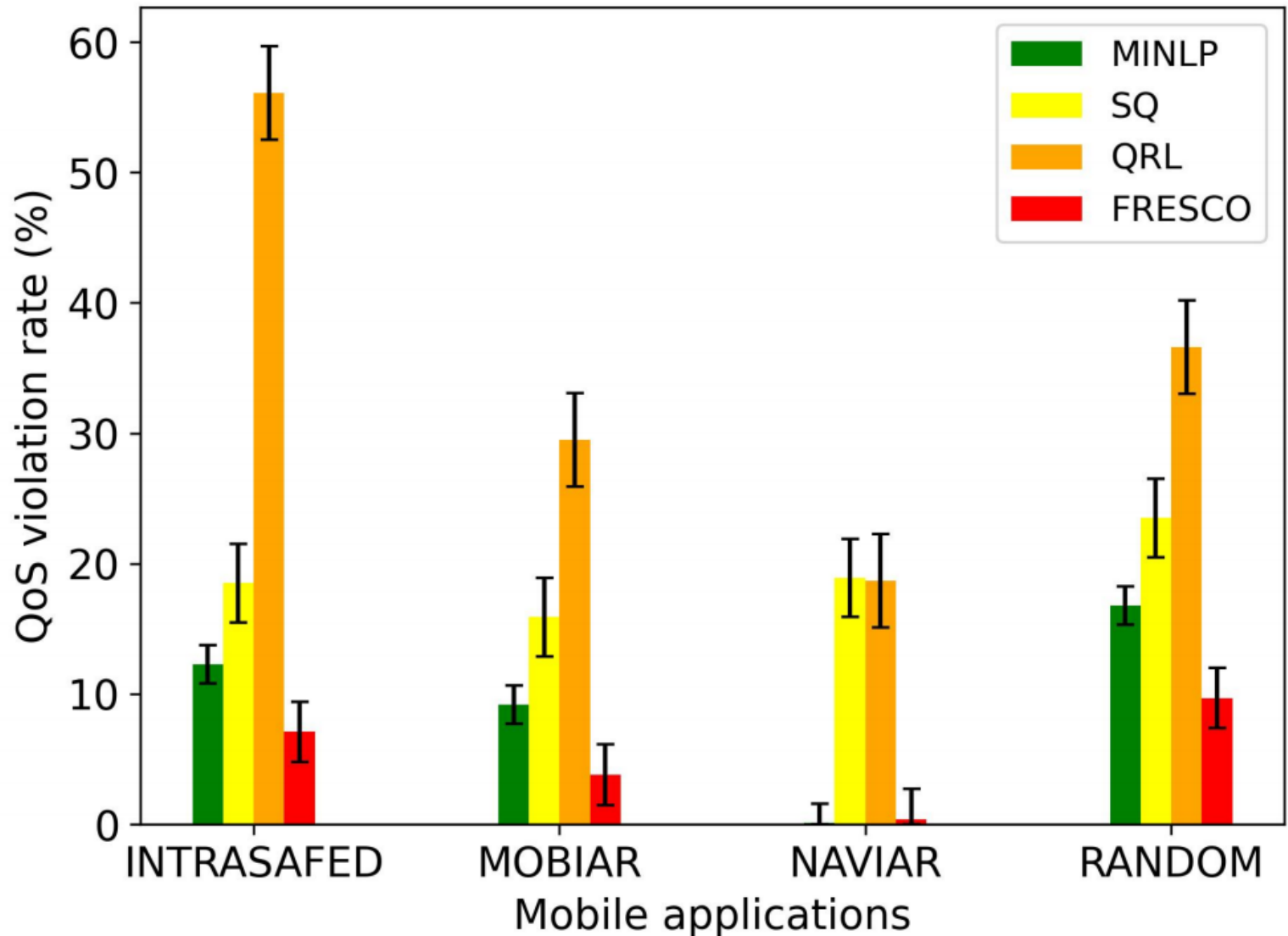}
    \caption{QoS violations}
    \label{fig:constraint-violations}
\end{figure}

Figure~\ref{fig:constraint-violations} illustrates QoS violation results. \revnote{R1\_C2}{QRL has the highest violation rate because of frequent offloading on local mobile and highly available clouds. Leading to fewer failures but frequent violations.} The next better-performing solution is SQ, with a violation rate between $18.9\%$ (in the NaviAR case) and $15.9\%$ (in the MobiAR case). MINLP shows better performance with violation rates of $12.3\%$, $9.2\%$, and $0.1\%$ in Intrasafed, MobiAR, and NaviAR, respectively. FRESCO has the lowest violation rates in Intrasafed and MobiAR use cases, with $7.1\%$ and $3.8\%$, and has a violation rate of $0.4\%$ with a standard deviation of $0.48\%$ in the NaviAR case, which is comparable with MINLP.

\subsubsection{\textbf{HSC overhead}}
HSC blockchain usage costs are expressed as gas consumption and are called Wei. The results are presented in Table~\ref{tab:block-overhead} for each function. Where range is expressed, it refers to executing from $1$ to $30$ offloading transactions, as multiple offloading transactions are typically executed for those functions. All HSC functions consume slightly above $21,000$ Wei, which is typical on the Ethereum~\cite{ethereumgasconsum}.

\begin{table}[t]
\centering
    \caption{Hybrid smart contract usage cost on Ethereum} \label{tab:block-overhead}
    \begin{tabular}{ | m{3.2cm} | m{4cm}|  } 
        \hline
        \textbf{Function name} & \textbf{Gas consumption (Wei)}\\
        \hline
        registerNode & $21,503$ Wei\\
        \hline
        unregisterNode & $21,204$ Wei\\
        \hline
        getNodeCount & $21,604$ Wei\\
        \hline
        getNode & $21,204$ Wei \\
        \hline
        updateNodeReputation & $21,638$-$29,984$ Wei\\
        \hline
        getReputationScore & $21,204$ Wei\\
        \hline
        resetReputation & $21,484$-$25,544$ Wei\\
        \hline
    \end{tabular}
\end{table}

\subsubsection{\textbf{Offloading decision overhead}}
\begin{figure}[h]
    \centering    
    \includegraphics[width=0.4\textwidth,height=13.5em]{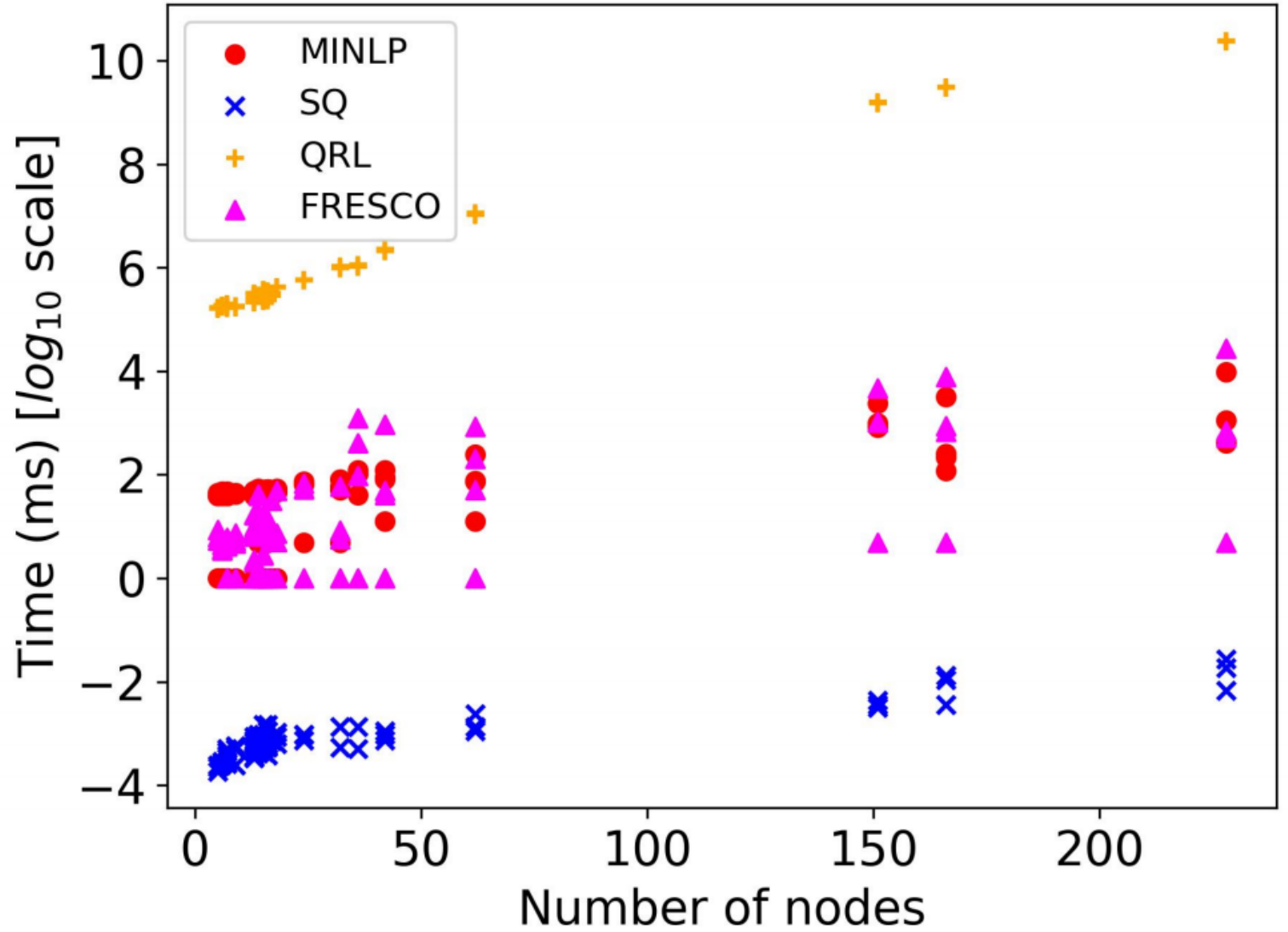}
    \caption{Offloading decision time in logarithmic scale}
    \label{fig:MINLP-overhead}
\end{figure}

Figure~\ref{fig:MINLP-overhead} illustrates offloading decision time overhead \revnote{R1\_C2}{ across different infrastructure sizes on a logarithmic scale measured on an AMD64 CPU 1.8GHz server}. SQ is the least complex algorithm, since selecting the first \textit{k} nodes and computing their estimated queue waiting time is relatively straightforward in comparison to other decision engines. The average decision time overhead is $0.048$ milliseconds. FRESCO and MINLP decision time overheads are $5.05$ milliseconds and $6.57$ milliseconds with standard deviations of $5.16$ milliseconds and $3.07$ milliseconds, respectively, making them comparable. \revnote{R1\_C2}{QRL has the highest overhead, which is an average of $1373.83$ milliseconds, due to the state space explosion when offloading on a larger number of nodes.} To summarize, FRESCO has a suitable decision time performance of $5.05$ milliseconds on average, which makes it a suitable candidate for latency-sensitive requirements.
 
\textbf{Summary:} FRESCO decreases average response time up to $7.86x$, and increases battery up to $5.4\%$ compared to baselines. It also achieves a low deadline violation rate of $0.4\%$ while maintaining competitive utilization costs. With approximately typical blockchain consumption ($\approx21,500$ Wei) and low average decision time overhead ($5.05$ milliseconds), FRESCO is suitable for offloading latency-sensitive applications.

\subsection{\rev{Sensitivity Analysis}}
\revnote{R1\_C1/R2\_C5}{We perform a sensitivity analysis of the FRESCO solution as part of our experiment with the goal to quantitatively assess robustness and adaptability in terms of multi-objective trade-off dynamics (latency vs.\ battery vs.\ cost) by adjusting hyperparameters (i.e., $\alpha$, $\beta$, $\gamma$). It serves as a guide for hyperparameter tuning to meet diverse user needs and application requirements. For instance, preferring costly but fast service for latency-sensitive applications, slower but more energy-efficient solutions for battery-constrained devices, or cost-saving options for budget-friendly enterprises. In the following sensitivity plots, we indicate with horizontal dash lines the best performance recorded during our experiment for better comparison, since the scale of objectives can vary greatly.}

\begin{figure*}[ht]
  \centering

  \begin{subfigure}[t]{0.33\textwidth}
    \includegraphics[width=\linewidth, height=12em]{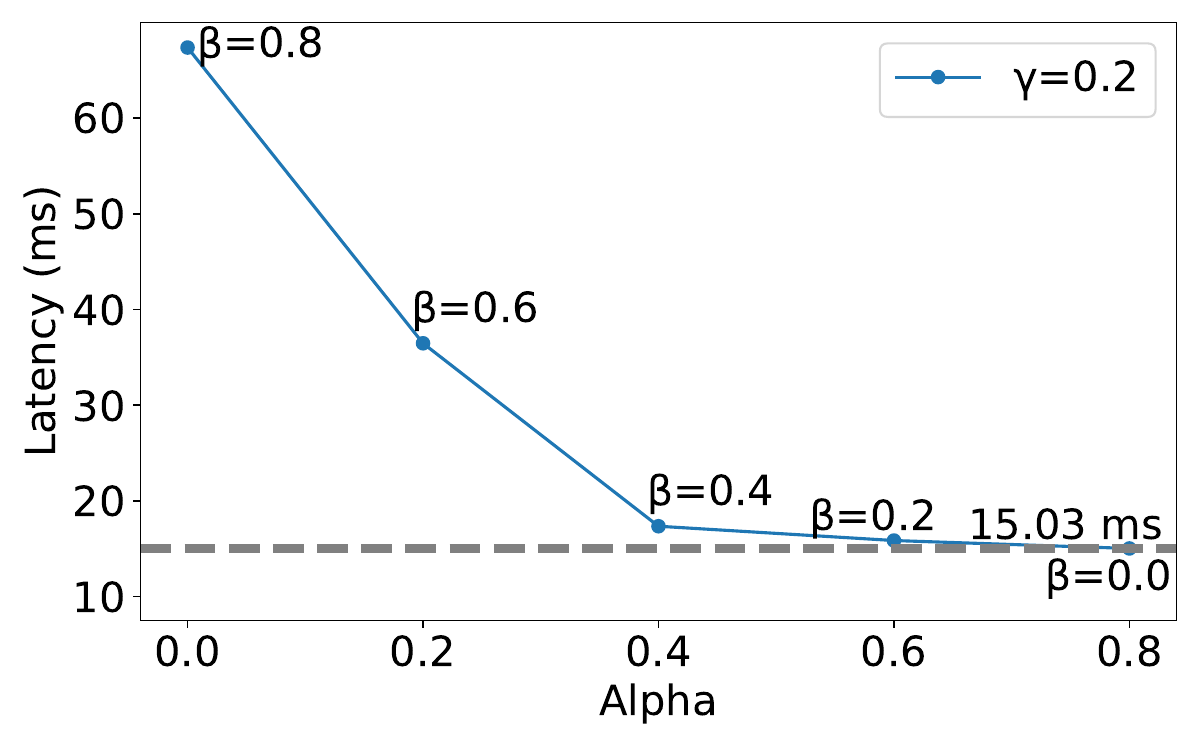}
    \caption{\rev{Latency ($\gamma$=0.2)}}
    \label{fig:latency_gamma_0.2}
  \end{subfigure}
  \begin{subfigure}[t]{0.33\textwidth}
    \includegraphics[width=\linewidth, height=12em]{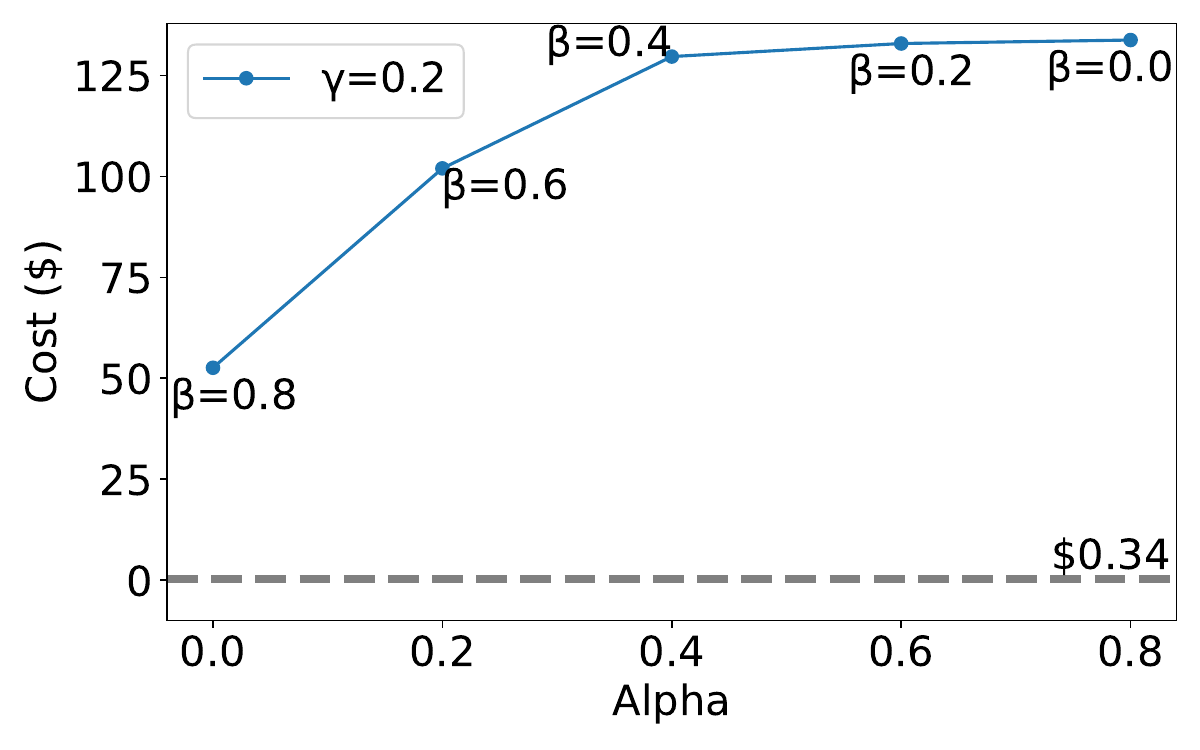}
    \caption{\rev{Cost ($\gamma$=0.2)}}
    \label{fig:cost_gamma_0.2}
  \end{subfigure}
  \begin{subfigure}[t]{0.32\textwidth}
    \includegraphics[width=\linewidth, height=12em]{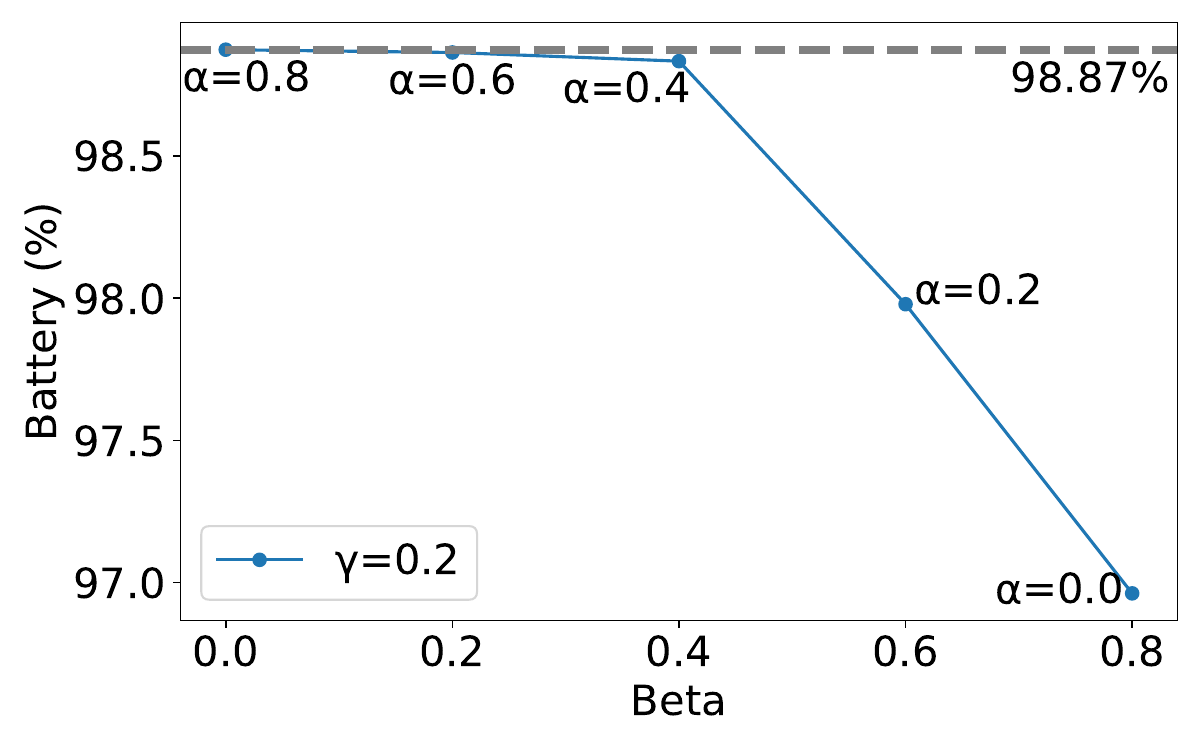}
    \caption{\rev{Battery ($\gamma$=0.2)}}
    \label{fig:battery_gamma_0.2}
  \end{subfigure}

  \begin{subfigure}[t]{0.33\textwidth}
    \includegraphics[width=\linewidth, height=12em]{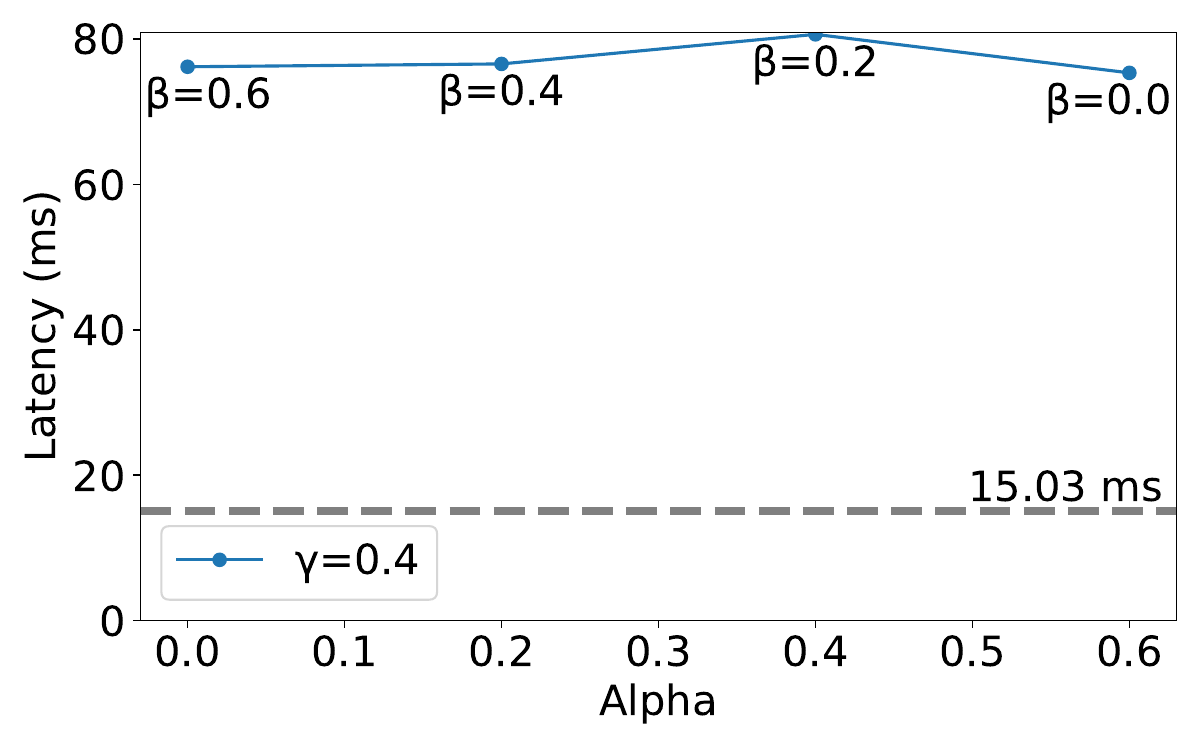}
    \caption{\rev{Latency ($\gamma$=0.4)}}
    \label{fig:latency_gamma_0.4}
  \end{subfigure}
  \begin{subfigure}[t]{0.33\textwidth}
    \includegraphics[width=\linewidth, height=12em]{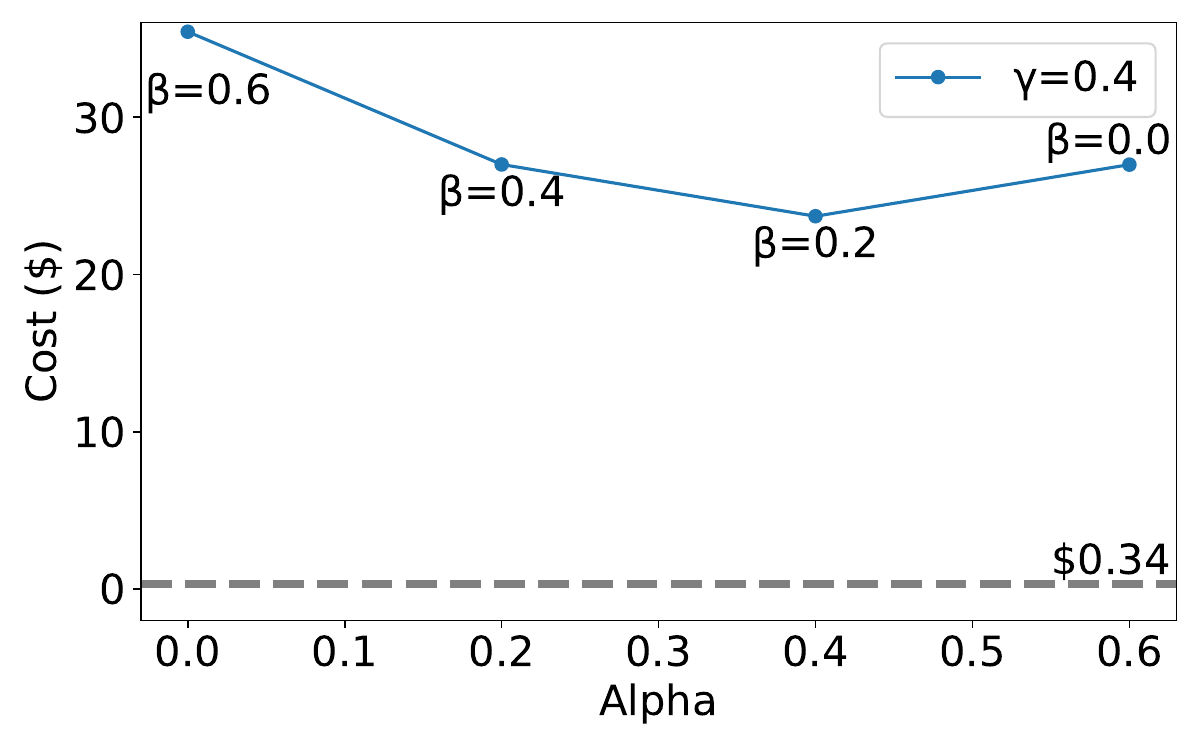}
    \caption{\rev{Cost ($\gamma$=0.4)}}
    \label{fig:cost_gamma_0.4}
  \end{subfigure}
  \begin{subfigure}[t]{0.32\textwidth}
    \includegraphics[width=\linewidth, height=12em]{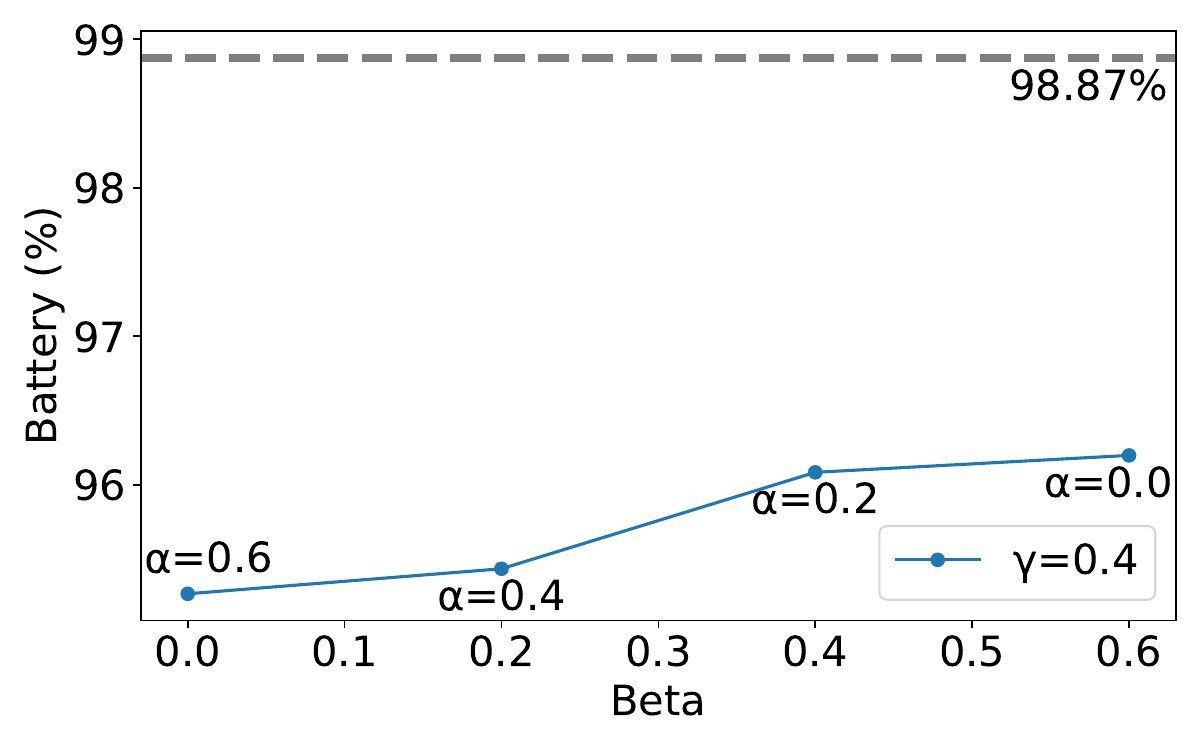}
    \caption{\rev{Battery ($\gamma$=0.4)}}
    \label{fig:battery_gamma_0.4}
  \end{subfigure}

  \begin{subfigure}[t]{0.33\textwidth}
    \includegraphics[width=\linewidth, height=12em]{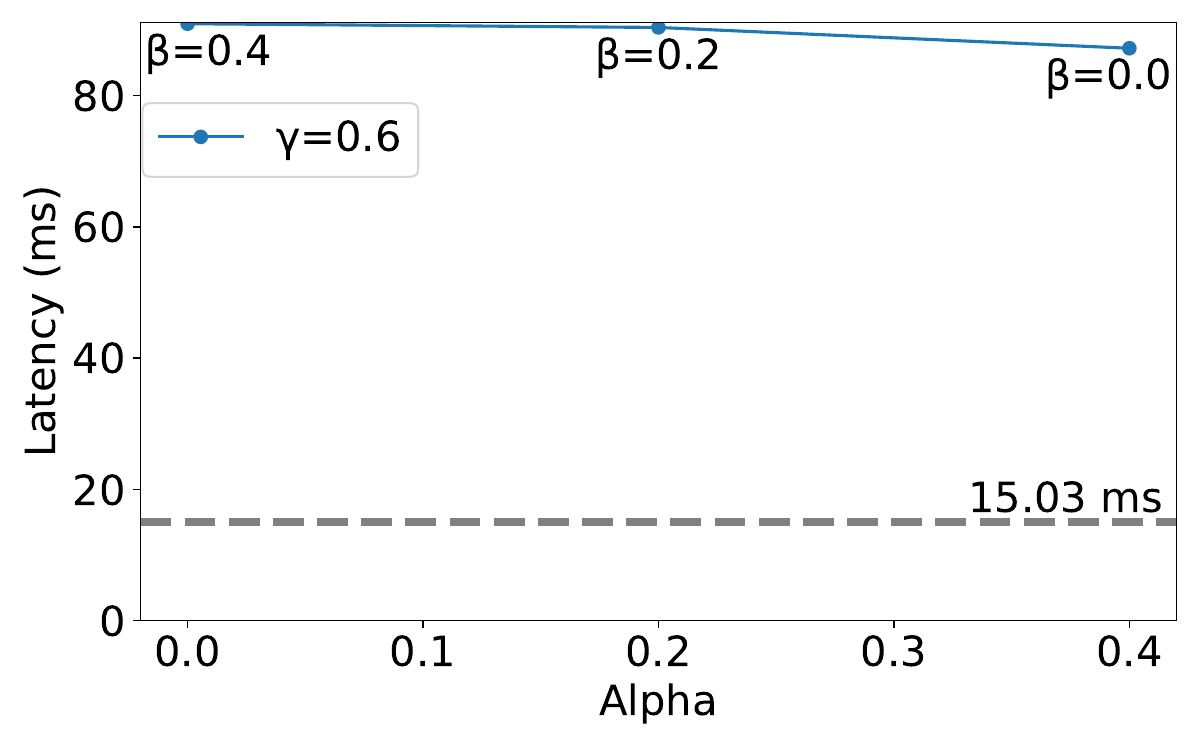}
    \caption{\rev{Latency ($\gamma$=0.6)}}
    \label{fig:latency_gamma_0.6}
  \end{subfigure}
  \begin{subfigure}[t]{0.33\textwidth}
    \includegraphics[width=\linewidth, height=12em]{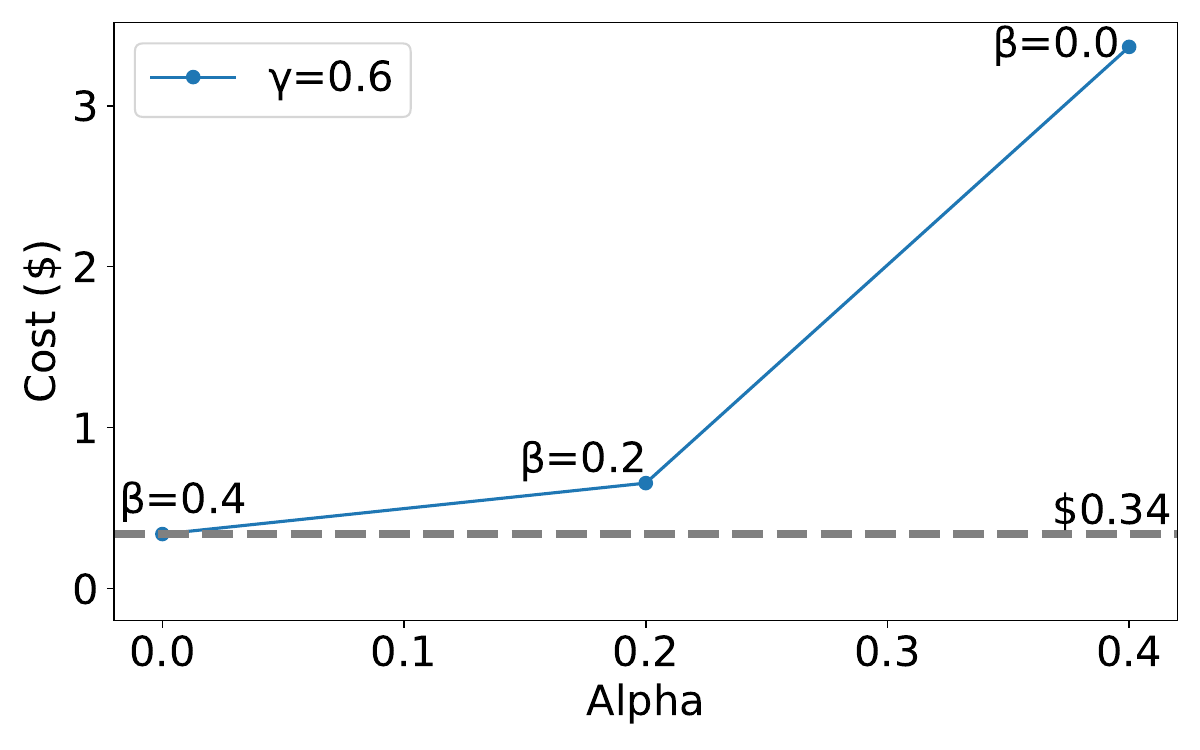}
    \caption{\rev{Cost ($\gamma$=0.6)}}
    \label{fig:cost_gamma_0.6}
  \end{subfigure}
  \begin{subfigure}[t]{0.32\textwidth}
    \includegraphics[width=\linewidth, height=12em]{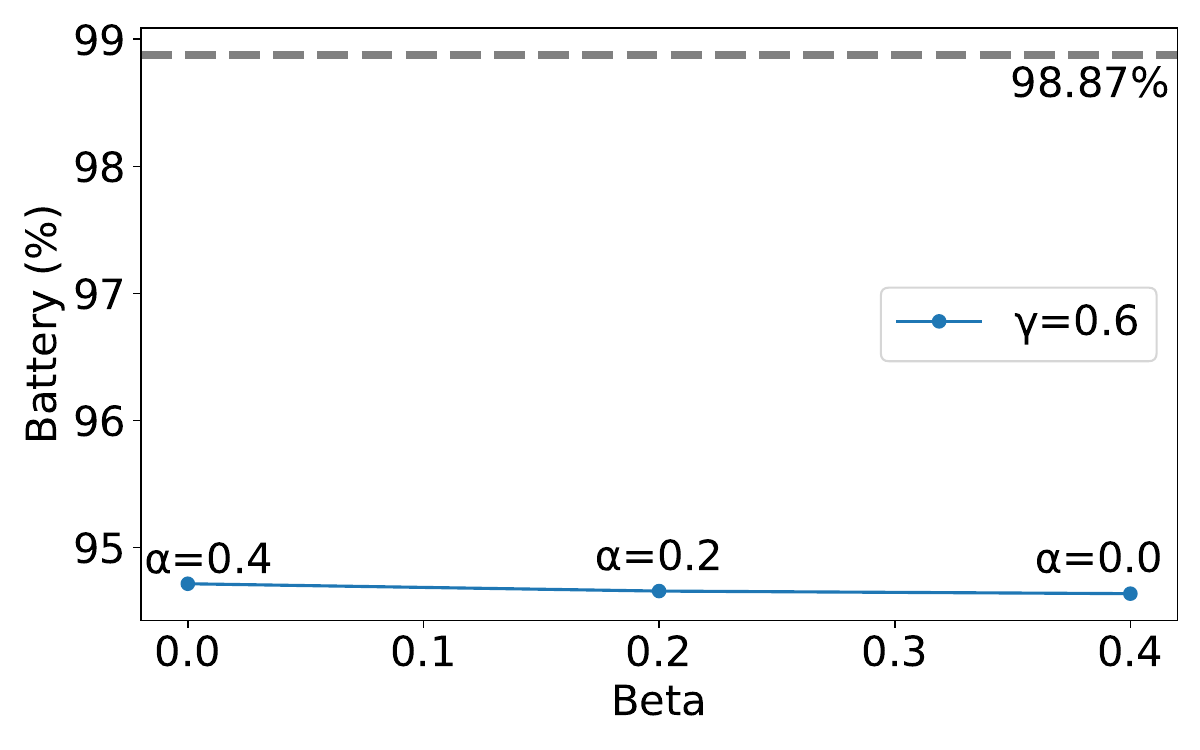}
    \caption{\rev{Battery ($\gamma$=0.6)}}
    \label{fig:battery_gamma_0.6}
  \end{subfigure}

  \caption{\rev{FRESCO sensitivity analysis with different weights. Best performance is marked with a horizontal dashed line.}}
  \label{fig:fresco_sensitivity_grid}
\end{figure*}

\rev{The first three subplots have fixed $\gamma$ values (i.e., weight for resource utilization cost) at 0.2, indicating a lower preference for cost savings but preferring faster and more energy-efficient solutions, such as latency-sensitive applications on battery-powered mobile devices. In the first subplot (Figure~\ref{fig:fresco_sensitivity_grid}\subref{fig:latency_gamma_0.2}), increasing $\alpha$ response time weight reduces latency exponentially from 67.35 seconds ($\alpha$ = 0, $\beta$ = 0.8) to 15.03 milliseconds ($\alpha$ = 0.8, $\beta$ = 0). In the second subplot, an increasing reliance on $\alpha$ shows an increasing cost trend where faster latency means utilizing edge servers more frequently, and thus higher costs, up to 133.78 monetary units. The battery lifetime in the third subplot shows stable and high capacity conservation, as long response time weight $\alpha = [0.6, 0.8]$ is dominant or balanced with battery lifetime importance $\alpha = 0.4, \beta = 0.4$. Tipping point is when the battery weight $\beta$ becomes more significant above $0.4$ at the expense of $\alpha$ response time, then the battery starts to reduce down to 97\%. Here, FRESCO offloads to a cheaper cloud rather than an expensive edge, but the longer cloud transmission latency drains the device battery more.}


\rev{The fourth to sixth subplots, where $\gamma=0.4$, target a more balanced parameter trade-off for general-purpose applications. The latency subplot (Figure~\ref{fig:fresco_sensitivity_grid}\subref{fig:latency_gamma_0.4}) shows stable but high latency values around 77 milliseconds per task execution. Solid $\gamma$ value prefers cheaper options like mobile and cloud, and thus higher latency. Cost subplot (Figure~\ref{fig:fresco_sensitivity_grid}\subref{fig:cost_gamma_0.4}) downscales cost between 23 and 35 monetary units compared to $\gamma = 0.2$, which are much higher. Battery subplot (Figure~\ref{fig:fresco_sensitivity_grid}\subref{fig:battery_gamma_0.4}) has a slight but noticeable increase in battery lifetime above 96\% when increasing $\beta$ up to 0.6.}

\rev{The last three subplots, where $\gamma = 0.6$, target enterprise users with budget-aware constraints. Subplot (Figures~\ref{fig:fresco_sensitivity_grid}\subref{fig:latency_gamma_0.6}) shows similar trend as (Figure~\ref{fig:fresco_sensitivity_grid}\subref{fig:latency_gamma_0.4}) while Figures~\ref{fig:fresco_sensitivity_grid}\subref{fig:cost_gamma_0.6} shows lowest cost mostly using low-cost cloud and no-cost mobile, and consequently battery lifetime is drained more due to longer cloud transmission latency in Figure~\ref{fig:fresco_sensitivity_grid}\subref{fig:battery_gamma_0.6}).}


\section{Discussion and Limitations} \label{sec:discussion}
Limitations of our experiment are the usage of a blockchain emulator and the selection of a single consensus mechanism. In the former, the real Ethereum requires tokens, which carry a financial cost and prevent us from collecting sufficient traces to strengthen the experimental evaluation. In the latter, the selected lightweight PoA consensus is used in real Ethereum \cite{ethereumblocktime}, has relatively shorter latencies compared to computationally intensive ones (e.g., Proof-of-Work), and does not test our solution on varied consensus latencies. Also, the choice to use public blockchain instead of private blockchain is that we targeted more open and public settings instead of being contained within certain organizations (e.g., enterprises).

Hyperparameter optimization of FRESCO could potentially increase the performance but also introduce overhead, thus endangering online applicability. User-defined hyperparameters enable flexibility to fit different needs (e.g., preferring cheaper but slower service over an expensive and faster one). The empirically obtained hyperparameters reflect a representative scenario where latency is preferred over other objectives, which is typical for latency-sensitive applications. Theoretically, without relying on user-defined values, mobile devices need to adjust them by incorporating lightweight data-driven optimization.

\section{Related Work} \label{sec:related}
\begin{table}[t]
    \caption{Overview of state-of-the-art literature}
    \label{tab:literature}
    \centering 
    \scriptsize
    \begin{tabular}{ |p{2.5cm}|p{0.7cm}|p{0.7cm}|p{1cm}|p{0.7cm}|p{0.7cm}|}
        \hline
        \textbf{Publication} & \textbf{OFF} & \textbf{(H)SC} & \textbf{BLOCK(-REP)} & \textbf{REL}\\
        \hline
        \cite{ren2021adaptive,zabihi2023reinforcement,zhang2021mobility} & \checkmark & \xmark & \xmark & \xmark\\        
        \cite{iqbal2021blockchain,iqbal2020blockchain,Ma2024BlockchainBasedTO,Shi2023DRLBasedVC,Zhou2021BlockchainbasedTS}
        & \checkmark & \xmark & \checkmark & \xmark \\
        \cite{zilic2022edge,liang2023reliability,liu2023toward}
        & \checkmark & \xmark & \xmark & \checkmark\\
        \cite{molina2018implementation,solaiman2021implementation}
        & \xmark & \checkmark & \xmark & \xmark \\
        \cite{sun2021rc,yu2020crowdr,kang2021optimizing,zhou2021blockchain} & \xmark & \xmark & \checkmark & \checkmark \\
        
        \cite{deng2020incentive} & \checkmark & \xmark & \checkmark & \xmark \\
        
        \rowcolor{Gray}
        \bf{This work} & \checkmark & \checkmark & \checkmark & \checkmark \\
        \hline
    \end{tabular}
\end{table}

Table ~\ref{tab:literature} compares FRESCO with the literature concerning offloading $OFF$, smart contracts $(H)SC$, blockchain-enabled or based-reputation $BLOCK(-REP)$, and reliability $REL$.

\revnote{R1\_C3}{Systematic literature review \cite {zabihi2023reinforcement} shows that applying deep reinforcement learning solutions (DRL) and their variants in edge offloading has become common recently due to their adaptability. Another offloading survey \cite{Dong2024TaskOS} outlines all recent mobile edge computing offloading solutions, including multi-objective methods, which include evolutionary (e.g., NSGA-II), mathematical (e.g., convex approximations), and metaheuristics (e.g, swarm intelligence) methods. Typically, these works exhibit centralized or semi-centralized solution deployments located on IIoT gateways, edge servers, and central orchestrators as an intermediary between mobile devices and remote servers. Since FRESCO targets scenarios with unreliable servers, these solution deployments are subject to a single point of failure. Moreover, it may require a model retraining when the environment changes drastically \cite {zhang2021mobility}}. 



Blockchain-enabled edge offloading approaches \cite{Ma2024BlockchainBasedTO,Shi2023DRLBasedVC,Zhou2021BlockchainbasedTS} enhance reliability and efficiency in mobile edge computing and vehicular edge networks, where blockchain and smart contracts make offloading more secure and trustworthy. \revnote{R1\_C3}{\cite{Ma2024BlockchainBasedTO} does not provide specifics about real-world smart contract implementation, while \cite{Shi2023DRLBasedVC} does not demonstrate practical prototype and \cite{Zhou2021BlockchainbasedTS} deploys an offloading framework as a smart contract, which is unrealistic since smart contracts prohibit nondeterministic and stochastic computations that conflict with the determinism requirement of blockchain consensus.} 

Blockchain-based reputation offloading was proposed for specific scenarios that require fast responses, like vehicular networks \cite{iqbal2020blockchain} and IIoT \cite{iqbal2021blockchain}. Solutions are applied in private environments (e.g., factories, enterprises) while ignoring the consensus overhead, and using reputation against malicious actors rather than for reliability. Also, other blockchain-based reputation approaches are applied for selecting trustworthy edge servers, ranging from IoT \cite{yu2020crowdr}, federated edge learning \cite{kang2021optimizing} to vehicular networks \cite{sun2021rc}. However, they did not target reliability in edge offloading in terms of edge failures and do not employ HSC, which forces them to compute reputation off-chain rather than on-chain, which can lead to potential risks (e.g., collusion) \cite{deng2020trust}.

Some edge offloading approaches \cite{peng2022reliability,liang2023reliability,liu2023toward} tried to solve reliability issues in terms of failure and recovery probabilistic models, or predict edge failures based on historical data \cite{zilic2022edge}. The aforementioned works did not prove or evaluate their solutions in distributed unreliable edge scenarios where the device moves and reacts to different environments.

Works \cite{molina2018implementation,solaiman2021implementation} implemented smart contracts on a hybrid blockchain architecture to reconcile conflicting objectives, such as trust on one side and performance on the other side. The works are not applied in the edge offloading context.


\textbf{Summary:} None of the works applied a blockchain-reputation system for enhancing reliability in the offloading context with formal rigor for latency-sensitive applications. FRESCO uniquely ensures trust for sensitive reputation information on-chain while allowing fast and formally verified performance for latency-sensitive applications off-chain.

\section{Conclusion} \label{sec:conclusion}




We investigated edge offloading of latency-sensitive mobile applications on the distributed unreliable edge. Edge offloading is formulated as a constraint optimization problem that balances between response time, battery, and utilization monetary cost objectives. Formulation incorporates critical QoS deadlines that have to be respected, and reputation scores to identify reliable edge servers based on past performance.

The FRESCO consists of a reputation state manager and a decision engine. The reputation state manager is implemented as a hybrid smart contract, which stores sensitive reputation scores on-chain against tampering, and enables an SMT-based decision engine to compute offloading decisions off-chain on reliable servers without being hindered by blockchain consensus. The presented solution balances reliability and performance where trust is required against tampering.

FRESCO was evaluated against baselines with simulated applications, a dynamic queueing workload, Skype availability traces, and large-scale infrastructure from the OpenCellID. We also discussed limitations like the blockchain emulator, hyperparameter optimization, and using a single consensus mechanism, which will be addressed in our future work.

\section*{Acknowledgments}
This work is partially funded by Josip Zilic's netidee scholarship by the Internet Foundation Austria.

\bibliographystyle{IEEEtran}
\bibliography{references}












\section*{Biography Section}
 



\vspace{-30pt}
\begin{IEEEbiographynophoto}{Josip Zilic}
is a Pre-Doctoral Researcher at the High-Performance Research Group at TU Wien. His research focuses on applying formal methods in edge offloading to ensure and guarantee performance for latency-sensitive and high-reliability mobile applications. In 2023, he received netidee scholarship from Internet Foundation Austria for his edge offloading doctoral thesis proposal.
\end{IEEEbiographynophoto}
\vspace{-30pt}
\begin{IEEEbiographynophoto}{Vincenzo De Maio}
received his PhD in 2016 at the University of Innsbruck, Austria. His research in the area of parallel and distributed systems comprises energy-aware Cloud computing and scheduling. Since 2025, he is a Lecturer in Distributed Systems at the University of Leicester. He authored different conferences and journal publications on energy efficiency and modeling for Cloud, Edge, and Quantum Computing.
\end{IEEEbiographynophoto}
\vspace{-30pt}
\begin{IEEEbiographynophoto}{Shashikant Illager}
 is an assistant professor at the Informatics Institute, University of Amsterdam, Netherlands. He is a member Multiscale Networked Systems research group. He works at the intersection of distributed systems, energy efficiency, and machine learning. His recent research explores the energy efficiency and performance optimization of data-intensive and distributed AI applications.
\end{IEEEbiographynophoto}
\vspace{-30pt}
\begin{IEEEbiographynophoto}{Ivona Brandic}
is a Professor at TU Wien. In 2015 she was awarded FWF START prize, the highest Austrian award for young researchers. She received her PhD degree in 2007 from Vienna University of Technology. In 2011 she received the Distinguished 
Young Scientist Award from the Vienna University of Technology for her project on the Holistic Energy Efficient Hybrid Clouds. Her main research interests are cloud computing, large-scale distributed systems, energy efficiency, QoS, and autonomic computing.
\end{IEEEbiographynophoto}

\vfill
\clearpage

\newcolumntype{B}[1]{>{\ifhighlighted\color{blue}\fi}p{#1}}

\newif\ifhighlighted
\highlightedfalse  



{
\appendices

\section{Full Simulation Table}
\begin{table}[!htbp]\centering
    \caption{Parameters} \label{tab:notation}
    \captionsetup{labelfont={color=blue}, textfont={color=blue}}  
    \renewcommand{\arraystretch}{0.25}
    \begin{tabular}{ |B{3.6cm}|B{2cm}|B{10cm}| }

        \hline
        & \textbf{Notation} & \textbf{Description} \\
        \hline
        \multirow{9}{*}{Application and servers}
            & $v$ & Server \\
            & $t$ & Application task \\
            & $A$ & Application as set of tasks \\
            & $N$ & Set of infrastructure servers \\
            & $O_{\tau}$ & Set of offloaded tasks at time instant $\tau$ \\
            & $O_{\leq\tau}$ & Set of offloaded tasks before time instant $\tau$ \\
            & $\delta_{in} (t)$ & Set of task $t$ input dependencies\\
            & $data(t)$ & Task data \\
        \hline
        \multirow{8}{*}{Battery and energy}
            & $E(v,t,m,h_m)$ & Total energy consumption of mobile device $m$ when offloading task $t$ on server $v$ through communication channel $h$\\
            & $BL(v,t,m,h_m)$ & Battery lifetime of mobile device $m$ after offloading task $t$ on server $v$ through device communication channel $h_m$ \\
            & $bcap$ & Battery capacity \\
            & $p_e (m)$ & Execution power on mobile device $m$\\
            & $p_c (h_m)$ & Communication power on mobile device on channel $h_m$\\
            &$C$ & Number of power state transitions\\
            &$T_{idle}$ & Time duration of idle power state\\
            &$Ch(h_m)$ & channel capacity of channel $h_m$\\
        \hline
        \multirow{4}{*}{Objective weights}
            & $\alpha$ & Response time objective weight \\
            & $\beta$ & Battery lifetime objective weight \\
            & $\gamma$ & Cost objective weight \\
            & $score$ & Multi-objective scoring \\
        \hline
        \multirow{4}{*}{Latency parameters}
            & $AP\tau$ & Overall application response time until time instant $\tau$ \\
            & $T_o(h,t)$ & Task offloading latency of task $t$ via channel $h$\\
            & $T_e(v,t)$ & Task execution latency of task $t$ on server $v$ \\
            & $T_d(h,t)$ & Task delivery latency of task $t$ via channel $h$\\
            & $T_c(h,t)$ & Task communication latency, which is generic both for offloading and delivery \\
        \hline
        \multirow{5}{*}{Cost paramters}
            & $PR(v,t)$ & Resource utilization cost of task $t$ on server $v$\\
            & $cost_r$ & Remote cloud cost \\
            & $cost_e$ & Edge penalty cost \\
            & $cost_{cores}$ & CPU price \\
            & $cost_{stor}$ & Storage price \\
        \hline
        \multirow{4}{*}{Reputation parameters}
            & $R_\tau(v,t,h)$ & Reputation score at time $\tau$ after task $t$ execution on server $v$ through communication channel $h$\\
            & $inc_\tau(v,t,h)$ & Task incentive after task $t$ executed on server $v$ through communication channel $h$ at time instant $\tau$\\
            & $\omega$ & Reputation update weight \\
        \hline
        \multirow{3}{*}{Queuing and load}
            & $\lambda_c(v)$ & Communication task arrival rate on server $v$\\
            & $\lambda_e(v)$ & Execution task arrival rate on server $v$\\
            & $U_c(h,t)$ & Communication queue utilization on channel $h$ when offloading task $t$\\
            & $MIPS(v)$ & Computational capacity of server $v$ in millions of instructions per second\\
            & $MI(t)$ & Instruction count of task $t$ \\
            & $\omega_e(v)$ & Execution waiting time on server $v$\\
            & $\omega_c(h,t)$ & Communication waiting time on channel $h$ when offloading task $t$\\
            & $\mu_e(v,t)$ & Execution service time on server $v$ when executing task $t$\\
            & $\mu_c(h,t)$ & Communication service time on channel $h$ when offloading o delivering task $t$\\
            & $bw_{avail}(h)$ & Available bandwidth on channel $h$\\
            & $bw_{total}(h)$ & Total bandwidth of channel $h$\\
            & $bw_{util}(h)$ & Utilized bandwidth on channel $h$ \\
        \hline
        \multirow{5}{*}{SMT constraints}
            & $stor(v)$ & Storage capacity on server $v$ \\
            & $cpu(v)$ & CPU capacity of server $v$ \\
            & $mem(v)$ & Memory capacity of server $v$ \\
            & $\delta_{in}(t)$ & Input dependencies of task $t$ \\            
            & $\nabla$ & Task time constraint \\
            & $D$ & Application deadline \\
            & $k$ & Top $k$ reliable server candidates for offloading\\
            & $rp$ & Reputation threshold \\
        \hline
    \end{tabular}
\end{table}
}


\end{document}